\documentclass[journal]{IEEEtran}
\IEEEoverridecommandlockouts
\usepackage{setspace}
\usepackage{cite}
\usepackage{amsmath,amssymb,amsfonts,mathtools,bm}
\usepackage{amsthm}
\usepackage{algorithm}
\usepackage{algpseudocode}
\usepackage{graphicx}
\usepackage{textcomp}
\usepackage{xcolor,float}
\usepackage{tabularx}
\usepackage{textcomp}
\usepackage{diagbox}
\usepackage{svg}
\usepackage{booktabs}
\usepackage{subfigure}
\usepackage{hyperref}

\usepackage{makecell}
\usepackage{amsthm}
\usepackage{sidecap}
\usepackage{microtype}
\usepackage{booktabs} 
\usepackage{multirow}
\usepackage{float}
\usepackage{adjustbox} 
\usepackage{amsmath}
\usepackage{amssymb}
\usepackage{colortbl}
\usepackage{makecell}
\usepackage{float}
\usepackage[table]{xcolor}
\usepackage{url}
\usepackage{balance}
\usepackage{bbding}
\usepackage{sidecap}
\usepackage{microtype}
\usepackage{booktabs} 
\usepackage{multirow}
\usepackage{array}
\usepackage{makecell}
\usepackage{amsmath}
\usepackage{amssymb}

\usepackage{tikz}
\newlength{\qualpanelwidth}
\newlength{\qualrowlabelwidth}
\usepackage{booktabs}
\usepackage{multirow}
\usepackage{xcolor}

\definecolor{bestred}{RGB}{185,30,30}
\definecolor{secondblue}{RGB}{30,70,175}
\definecolor{gaincyan}{RGB}{0,145,175}
\definecolor{gainorange}{RGB}{225,105,20}

\newcommand{\best}[1]{\textcolor{bestred}{\textbf{#1}}}
\newcommand{\second}[1]{\textcolor{secondblue}{\underline{#1}}}
\newcommand{\gainlow}[1]{\textcolor{gaincyan}{\ensuremath{\downarrow}\,#1}}
\newcommand{\gainhigh}[1]{\textcolor{gainorange}{\ensuremath{\uparrow}\,#1}}

\usepackage{changes} 

\definecolor{lime}{HTML}{A6CE39}
\DeclareRobustCommand{\orcidicon}{%
    \begin{tikzpicture}
    \draw[lime, fill=lime] (0,0) 
    circle [radius=0.16] 
    node[white] {{\fontfamily{qag}\selectfont \tiny ID}};    \draw[white, fill=white] (-0.0625,0.095) 
    circle [radius=0.007];    \end{tikzpicture}
    \hspace{-2mm}}
\foreach \x in {A, ..., Z}{%
    \expandafter\xdef\csname orcid\x\endcsname{\noexpand\href{https://orcid.org/\csname orcidauthor\x\endcsname}{\noexpand\orcidicon}}
    }

\usepackage{orcidlink}  

\newcommand{\orcidauthorA}{0009-0009-3708-9196} 
\newcommand{\orcidauthorB}{0000-0003-1439-4875} 
\newcommand{\orcidauthorC}{0000-0001-7907-2071} 

\allowdisplaybreaks
\renewcommand{\baselinestretch}{1}

\begin{document}

\title{BeamRMX: Radiation-Pattern-Driven Learning for Generalizable Beam Radio Map Prediction and Beam Management}

\author{
Yue Zhang~\orcidlink{\orcidauthorA},~\IEEEmembership{Student Member, IEEE},
Xiucheng Wang~\orcidlink{\orcidauthorB},~\IEEEmembership{Graduate Student Member, IEEE},
Wenshuo Chen,~\IEEEmembership{Student Member, IEEE},
and Nan Cheng~\orcidlink{\orcidauthorC},~\IEEEmembership{Senior Member, IEEE}

\thanks{This work was supported by the National Key Research and Development Program of China (2024YFB907500).}

\thanks{Yue Zhang, Xiucheng Wang, and Nan Cheng are with  the State Key Laboratory of ISN and the School of Telecommunications Engineering, Xidian University, Xi'an 710071, China (e-mail: 25011210995@stu.xidian.edu.cn; xcwang\_1@stu.xidian.edu.cn; dr.nan.cheng@ieee.org). Corresponding author: Nan Cheng.}

\thanks{Wenshuo Chen is with The Hong Kong University of Science and Technology (Guangzhou), Guangzhou 511453, China.}
}

    \maketitle

\IEEEdisplaynontitleabstractindextext

\IEEEpeerreviewmaketitle
\begin{abstract}
The evolution toward sixth-generation (6G) wireless networks is driving larger antenna arrays and highly directional multi-beam transmission, making accurate knowledge of beam-dependent spatial coverage important for beam management and environment-aware network operation. Radio maps (RMs) provide such a representation, yet conventional RM prediction assumes omnidirectional or transmitter-level radiation. In beamformed multiple-input multiple-output (MIMO) systems, one propagation scene instead gives rise to many configuration-dependent beam radio maps (BeamRMs), creating challenges in beam representation and generalization. Existing methods either condition prediction on beam descriptors or use beam maps as auxiliary inputs to generic architectures. We propose BeamRMX, which, to the best of our knowledge, is the first dedicated framework to treat the spatial radiation pattern as the primary BeamRM query and learn how scene geometry transforms it into the received power field. XBase learns multiscale interactions between the radiation query and scene geometry, while an optional Evidence Adapter uses a few cross-configuration BeamRMs from the same scene. Matched-domain and zero-shot experiments show consistent gains over deterministic and diffusion baselines, including mean absolute error reductions of 26.1\% on unseen scenes and 47.8\% on an unseen configuration. Cross-configuration evidence further improves reconstruction and intra-sector beam refinement. Code is available at \url{https://github.com/ZY021023/BeamRMX-code}.
\end{abstract}


\section{Introduction}
\label{sec:introduction}

\begin{figure*}[!t]
    \centering
    \includegraphics[width=\textwidth]{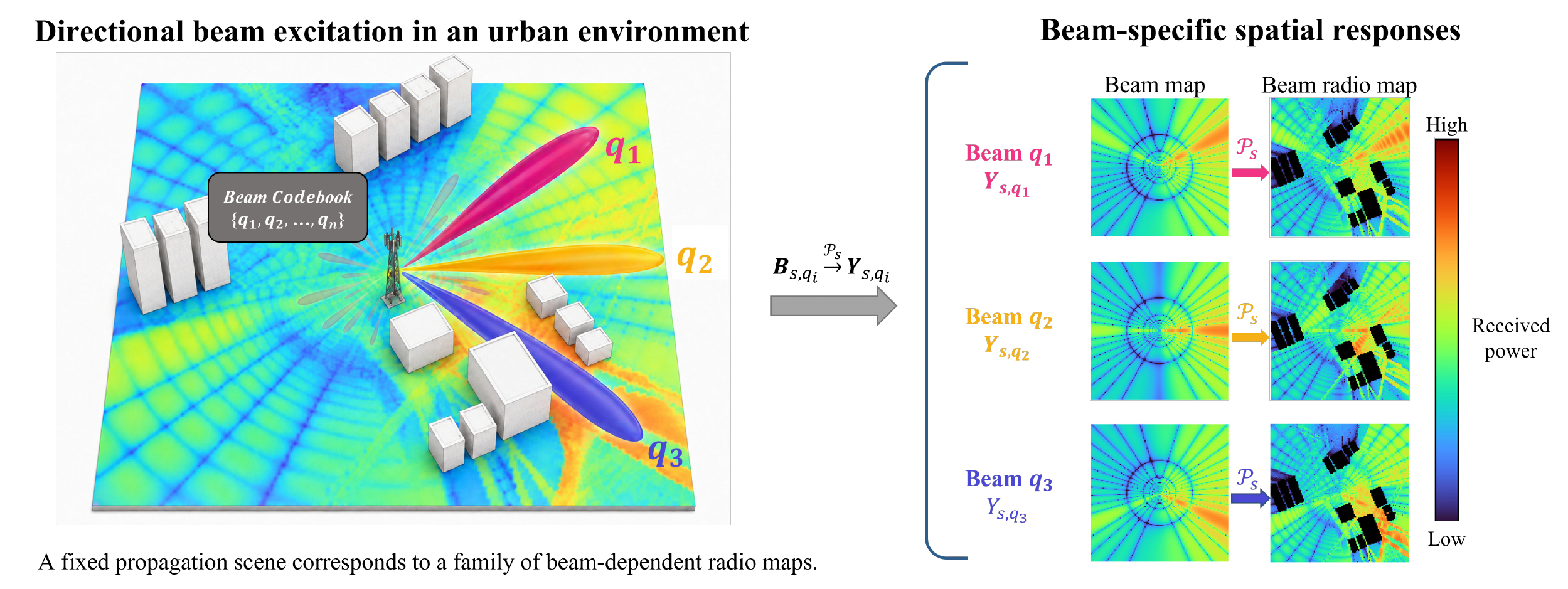}
    \caption{BeamRM concept: a common propagation scene transforms different beamforming-induced radiation patterns into distinct received power maps.}
    \label{fig:beamrm_concept}
\end{figure*}

The evolution toward sixth-generation (6G) wireless networks is driving the use of wider frequency bands, larger antenna arrays, and increasingly directional multi-beam transmission \cite{wang2024tutorial,dang2020should,chengnan6g,10638123,cheng2019space,heath2016mmwave,giordani2019beammanagement}. While beamforming provides substantial array gain and spatial selectivity, it also makes the resulting radio environment strongly dependent on the active beam and radio configuration. A beamforming state determines the transmitted radiation pattern but does not directly reveal the received power distribution after interaction with the propagation environment. Accurate knowledge of such beam-dependent spatial radio conditions is therefore important not only for beam management but also for coverage assessment, network planning, and other environment-aware wireless functions.

Radio maps (RMs) provide a natural representation of spatial radio conditions by associating geographical locations with received signal characteristics \cite{Alkhateeb2019DeepMIMOAG,zeng2024tutorial,Hoydis2022SionnaAO}. Existing RM prediction methods have achieved substantial progress, but most assume omnidirectional or transmitter-level radiation, under which a propagation scene is associated with a largely fixed spatial response \cite{levie2021radiounet,wang2024radiodiff,li2026u6gxlmimoradiomapprediction,radiodiff3ddata}. This assumption no longer holds in beamformed multiple-input multiple-output (MIMO) systems. Even for the same transmitter and environment, different carrier frequencies, array configurations, and beamforming states can produce substantially different received power fields, as illustrated in Fig.~\ref{fig:beamrm_concept}. We refer to these configuration-dependent spatial responses as beam radio maps (BeamRMs).

Despite substantial progress in learning-based RM prediction, dedicated work on BeamRM prediction remains limited. Recent studies have begun to model beam-dependent spatial responses, but how the target beam should be represented to the predictor remains a central issue. BeamCKM performs beam-aware channel-map construction through codebook-dependent beam representations, while BeamCKMDiff extends the conditioning variable to continuous beamforming vectors \cite{wang2025beamckm,zhao2026beamckmdiff}. The former distinguishes radiation states through predefined beam identities, whereas the latter provides a continuous description of the beamforming state. In both cases, however, the beam is represented primarily in the parameter domain, leaving the network to infer how the corresponding radiation pattern is distributed over the prediction region. This dependence on beam-specific parameterization becomes increasingly restrictive when prediction must extend across beam states, codebooks, or array configurations that differ from those seen during training.

The recently introduced beam map provides a more explicit representation by analytically projecting the radiation pattern associated with each radio configuration onto the prediction region \cite{li2026u6gxlmimoradiomapprediction}. In the original extremely large-scale MIMO (XL-MIMO) benchmark, beam maps are supplied as physics-informed spatial features to existing predictors. BeamRMX takes a further step by using this spatial radiation representation as the primary query for BeamRM prediction rather than as an auxiliary input. Beam indices, beamforming vectors, codebooks, and array configurations are thereby expressed in a common spatial domain before propagation is learned. This representation depends on the resulting spatial radiation pattern rather than on the particular parameters used to generate it, providing a more suitable basis for prediction across heterogeneous beam and array configurations.

Making the beam query spatial also exposes a second gap in existing methods. The spatial radiation pattern specifies how power is launched toward the environment, whereas scene geometry determines how that power is blocked, attenuated, and redistributed. Existing benchmarks using beam maps mainly concatenate these two information sources and rely on generic prediction architectures to learn their interaction implicitly \cite{li2026u6gxlmimoradiomapprediction}. Thus, although the beam map reduces the representation gap, a dedicated learning method is still needed to model how the propagation scene transforms the specified radiation pattern into the received power field. BeamRMX addresses this gap through XBase, its complete query-only predictor, which encodes the spatial radiation query and scene geometry separately and learns their multiscale interaction. To the best of our knowledge, XBase is the first architecture designed specifically for this formulation.

Query-only prediction is essential when no radio observation is available for a new deployment. In practice, however, a new target configuration rarely appears in a completely unobserved environment. As arrays grow larger and beams become narrower, the number of beam states required to characterize a deployment increases substantially, making exhaustive per-beam measurement or repeated high-fidelity simulation increasingly costly. Meanwhile, wireless networks evolve incrementally rather than being rebuilt from scratch. Before introducing a high-frequency, large-array, or narrow-beam configuration, the same area may already have been characterized under legacy frequency bands, smaller arrays, or wider-beam configurations. Measurements, simulations, and historical BeamRMs accumulated during earlier deployment stages therefore constitute a potentially valuable source of information for predicting the new configuration.

Source BeamRMs from the target scene contain realized information about scene-dependent blockage, attenuation, and multipath. However, they are not sparse observations of the target BeamRM because each source map corresponds to a different spatial excitation and cannot be directly copied or interpolated into the target field. The source-assisted problem is therefore to extract propagation information that is shared across configurations and use it to reduce the uncertainty of query-only prediction without observing the target BeamRM or updating model parameters at inference time.

BeamRMX addresses this setting with an optional Evidence Adapter. When source BeamRMs are unavailable, XBase operates as a complete query-only predictor. When a few source BeamRMs from the same scene are available, the adapter interprets them together with their corresponding source excitations and refines the frozen XBase prediction using the transferable cross-configuration evidence. The target configuration is excluded from the support set, and no target BeamRM observation or parameter update is required during inference. The main contributions of this paper are summarized as follows.
\begin{enumerate}
    \item A spatial-excitation-conditioned formulation of BeamRM prediction: Building on the beam map representation, we elevate spatial excitation from an auxiliary array-conditioning feature to the primary representation of a BeamRM query. This places heterogeneous beam and configuration states in a common spatial query domain and reformulates BeamRM prediction as learning how a propagation scene transforms a specified excitation field into the realized received power field, rather than requiring the predictor to infer the spatial radiation implicitly from configuration-dependent descriptors. Because the formulation depends on the spatial radiation itself rather than its particular parameterization, it can, in principle, extend beyond fixed codebooks and array beamforming to other spatially representable radiating-source patterns.

    \item The first dedicated architecture for spatial-excitation-conditioned BeamRM learning: To the best of our knowledge, XBase is the first dedicated learning architecture explicitly designed around this excitation-to-field formulation. It maintains distinct representations of spatial excitation and scene geometry and learns their structured interaction rather than relying on generic beam conditioning or input concatenation. The proposed method is systematically evaluated under matched-domain, scene-disjoint, and configuration-disjoint protocols to examine both learning effectiveness and query-only generalization.

    \item Source-assisted BeamRM prediction for incremental network evolution: We introduce an Evidence Adapter that exploits a few same-scene BeamRMs from legacy or lower-complexity radio configurations as cross-configuration propagation evidence for an unobserved target BeamRM. The adapter refines the frozen query-only prediction while preserving an exact query-only path and requiring neither target BeamRM observations nor test-time parameter updates. Evaluation under the strict-small protocol demonstrates that limited historical radio information can substantially improve BeamRM prediction in unseen scenes and enhance downstream intra-sector beam refinement.
\end{enumerate}
The remainder of this article is organized as follows. Section~II reviews related work, and Section~III formulates the query-only and source-assisted BeamRM prediction problems. Section~IV presents the proposed BeamRMX framework, while Section~V reports the experimental results and discussion. Finally, Section~VI concludes the article.

\section{Related Work}
\label{sec:related_work}

This section reviews learning-based radio map prediction, beam-aware representation, and generalization with auxiliary radio evidence.

\subsection{Learning-Based Radio Map Prediction}

Representative approaches include geometry-conditioned encoder--decoder networks, adversarial models, graph neural networks and diffusion-based predictors
\cite{levie2021radiounet,zhang2023rme,li2024radiogat,wang2024radiodiff}.
Diffusion-based RM prediction builds on denoising diffusion probabilistic modeling \cite{ho2020ddpm}, while recent physics-informed variants incorporate propagation-related constraints into generative reconstruction
\cite{radiodiffk2}.
Propagation-aware learning has also been explored by explicitly exploiting environmental semantics and geometry-dependent propagation structure
\cite{liu2023uavmap}.
Public datasets and benchmark challenges have further promoted standardized evaluation of learning-based radio map prediction
\cite{yapar2023first}.

These methods establish learning-based prediction as an efficient alternative to repeated measurement or ray tracing. However, most of these methods target transmitter-level radio maps under omnidirectional or weakly directional assumptions. Their primary task is therefore to learn an environment-to-coverage mapping, while the configuration-dependent spatial excitation produced by a beam is either omitted or represented only through auxiliary parameters. Direct application to BeamRM prediction consequently leaves a substantial part of the beam-dependent radiation behavior to be inferred from training data.

\subsection{Beam Representation for Beam-Aware Radio Map Prediction}
More broadly, channel knowledge maps (CKMs) provide site-specific representations of channel-related information for environment-aware communications \cite{zeng2021toward}. Directional information has been incorporated into radio map and channel map predictors through antenna-pattern projections, beam identities, and configuration descriptors. Element-pattern or directional-gain projections expose spatial directivity more explicitly than isotropic transmitter models \cite{jaensch2026aerial}, but generally do not capture the joint effects of array geometry, coherent beamforming, and carrier frequency.

BeamCKM represents different radiation states through codebook-dependent beam conditioning \cite{wang2025beamckm}. Such a representation distinguishes beams within a predefined family but does not explicitly describe how each beam illuminates the prediction region. BeamCKMDiff further replaces discrete beam identities with continuous beamforming-vector conditioning \cite{zhao2026beamckmdiff}, providing a richer description of the radiation state. Nevertheless, the beamforming vector remains a global descriptor, and the predictor must still infer how the underlying array and beamforming parameters translate into a spatial radiation pattern. These approaches therefore represent the beam primarily in the parameter domain rather than directly in the spatial domain where propagation occurs.

The beam map representation addresses this gap by analytically converting configuration-dependent radiation into a spatial feature over the prediction region \cite{li2026u6gxlmimoradiomapprediction}. It provides an explicit grid-aligned radiation prior before environment-dependent blockage and multipath are introduced. The original multi-configuration benchmark incorporates beam maps into existing predictors mainly through feature concatenation and demonstrates clear advantages over scalar configuration encoding, including under cross-configuration and cross-environment evaluation.

These studies mainly establish what directional information should be supplied to a BeamRM predictor. They do not fully address how that information should be processed. Existing models that incorporate beam maps largely retain generic image-to-image architectures and place spatial excitation and environmental information in a common feature stream, although the former specifies the radiation presented to the environment and the latter determines how that radiation is transformed. This leaves a distinct methodological gap between spatial beam representation and dedicated learning of the scene-conditioned excitation-to-field transformation.

\subsection{Generalization and Radio-Evidence-Assisted Prediction}
BeamRM generalization involves two complementary distribution shifts. Scene-disjoint evaluation tests transfer to unseen geographical layouts, whereas cross-configuration evaluation changes the frequency, antenna, radiation pattern, or beam condition presented to the environment. Previous studies have examined cross-environment prediction, cross-frequency transfer, antenna-pattern variation, and cross-beam generalization under different protocols \cite{li2026u6gxlmimoradiomapprediction,li2024radiogat,wang2025beamckm,zhao2026beamckmdiff}. A model that transfers across scenes does not necessarily generalize to unseen radiation conditions, and vice versa.

Auxiliary radio information has also been used to improve radio map construction. Sparse reconstruction methods assume that a subset of the target map is directly observed
\cite{teganya2022deepcompletion,qiu2024channel,zhang2023rme,sun2024matrix,li2024radiogat,radiodiffinverse}, while few-shot transfer and test-time adaptation specialize a predictor using limited target-domain information \cite{wang2026radiodifffs,sun2025radiopit}. BeamCKM further introduces M3ChanNet, which exploits environmental profiles and real-time multi-beam observations as side information for beam-aware map construction \cite{wang2025beamckm}. These approaches demonstrate the value of incorporating additional radio information, but differ from the source-assisted setting considered here. Our support consists of a few BeamRMs from other radio configurations in the same scene, while no same-scene BeamRM from the target configuration is observed.

Sparse target observations and cross-configuration source BeamRMs may both provide radio evidence, but their semantics are fundamentally different. The former directly samples the target field; the latter describes how the same environment has transformed other spatial excitations. Likewise, unlike few-shot or test-time parameter adaptation, the source-assisted setting considered here uses such cross-configuration responses as feed-forward evidence without updating model parameters. A suitable predictor must therefore preserve a reliable query-only path, extract only transferable scene information from source maps, and avoid treating source responses as direct substitutes for the target BeamRM.

Prior work has separately advanced geometry-based RM prediction, beam representation, and the use of auxiliary radio information. Generalizable BeamRM prediction requires these elements to be combined: heterogeneous radiation conditions need a common spatial representation, the scene-dependent transformation of that excitation must be learned, and source BeamRMs must be exploited without observing or optimizing on the target BeamRM. These requirements motivate the formulation considered here.
\section{Problem Formulation}
\label{sec:problem_formulation}

We formulate BeamRM prediction by separating three information roles: the target spatial excitation, the propagation scene, and optional same-scene radio evidence from other configurations. The first two define query-only prediction, whereas the third enables source-assisted prediction.

\subsection{BeamRM Prediction Conditioned on Spatial Excitation}
\label{subsec:spatial_excitation_formulation}

Consider a propagation scene $s\in\mathcal{S}$ represented over an $H\times W$ spatial region by
\begin{equation}
    \mathbf{G}_{s}
    \in
    \mathbb{R}^{C_g\times H\times W},
    \label{eq:scene_representation}
\end{equation}
where $\mathbf{G}_{s}$ contains the environmental information available to the predictor. Let $\Omega_s$ denote the corresponding spatial evaluation grid, including its coordinate registration relative to the transmitter.

A target radiation query is denoted by $q=(c,\rho)$, where $c$ specifies the underlying radio configuration, such as carrier frequency, array architecture, and antenna characteristics, while $\rho$ specifies the directional radiation state, such as the beamforming or steering state. Following the beam map formulation in \cite{li2026u6gxlmimoradiomapprediction}, its raw beam map over $\Omega_s$ is deterministically computed as
\begin{equation}
    \mathbf{B}_{\Omega_s,q}
    =
    \Psi_{\mathrm{beam}}(q;\Omega_s)
    \in
    \mathbb{R}^{H\times W}.
    \label{eq:beam_map}
\end{equation}
The subscript $s$ reflects only the evaluation grid and its transmitter-relative registration, rather than scene-dependent propagation. Once $q$ and $\Omega_s$ are fixed, the beam map contains no scene-dependent blockage or multipath information. We write $\mathbf{B}_{s,q}\equiv\mathbf{B}_{\Omega_s,q}$ hereafter.

The model-ready spatial excitation is constructed as
\begin{equation}
\mathbf{E}_{s,q}
=
\phi_{\mathrm{exc}}
\left(
\mathbf{B}_{s,q};\Omega_s
\right)
\in
\mathbb{R}^{C_e\times H\times W},
\end{equation}
where $\phi_{\mathrm{exc}}(\cdot)$ combines the beam map with transmitter-relative spatial cues. Thus, $\mathbf{E}_{s,q}$ serves as the spatial representation of the target radiation state. In the query-only path, the radio query $q$ affects the prediction through the beam map $\mathbf{B}_{s,q}$, and no separate query embedding is supplied to XBase.

The target BeamRM and its valid-region mask are denoted by
\begin{equation}
    \mathbf{Y}_{s,q}
    \in
    \mathbb{R}^{H\times W},
    \qquad
    \mathbf{M}_{s,q}
    \in
    \{0,1\}^{H\times W},
    \label{eq:target_beamrm}
\end{equation}
where $\mathbf{Y}_{s,q}$ is the corresponding received power field and $\mathbf{M}_{s,q}$ identifies valid pixels for supervision and evaluation. The mask itself is not treated as target-side radio evidence.

For a fixed scene, different radiation queries induce different spatial excitations and consequently different BeamRMs. We represent the underlying propagation process as
\begin{equation}
    \mathbf{Y}_{s,q}
    =
    \mathcal{T}_{s}(
        \mathbf{E}_{s,q}),
    \label{eq:excitation_to_field}
\end{equation}
where $\mathcal{T}_{s}$ denotes the unknown scene-conditioned propagation transformation. BeamRM prediction is therefore formulated as learning how a scene transforms a specified spatial excitation into the corresponding received power field, rather than directly associating a configuration descriptor with an output map.

\subsection{Query-Only Generalization Regimes}
\label{subsec:query_only_formulation}

Without same-scene radio evidence, prediction follows
\begin{equation}
    \widehat{\mathbf{Y}}^{\,0}_{s,q}
    =
    F^{0}_{\theta}
    \left(
        \mathbf{G}_{s},
        \mathbf{E}_{s,q}
    \right),
    \label{eq:query_only}
\end{equation}
where the superscript $0$ denotes query-only prediction.

We consider three complementary regimes. The \emph{matched} regime evaluates learning effectiveness when training and test samples follow the same scene--configuration distribution. The two zero-shot regimes satisfy
\begin{equation}
    \begin{aligned}
        \text{Scene-ZS:}\quad&
        \mathcal{S}_{\mathrm{tr}}
        \cap
        \mathcal{S}_{\mathrm{te}}
        =
        \varnothing,
        \\
        \text{Config-ZS:}\quad&
        \mathcal{C}_{\mathrm{tr}}
        \cap
        \mathcal{C}_{\mathrm{te}}
        =
        \varnothing,
    \end{aligned}
    \label{eq:zero_shot_regimes}
\end{equation}
where $\mathcal{S}$ and $\mathcal{C}$ denote scene and radio-configuration sets, respectively. Scene-ZS tests transfer to unseen propagation environments, whereas Config-ZS tests transfer to radio configurations entirely absent from training, including all of their directional states. The two regimes are evaluated independently to isolate the corresponding distribution shifts. No same-scene BeamRM is available at inference time in any query-only regime.

\subsection{Strict-Small Source-Assisted Prediction}
\label{subsec:source_assisted_formulation}

We further consider an unseen target scene $s$ for which a small number of BeamRMs are available under other radio configurations. Let $q_i=(c_i,\rho_i)$ denote the $i$th source query in the same scene. The $K$-shot support set is
\begin{equation}
    \mathcal{Z}^{K}_{s,q}
    =
    \left\{
        \left(
            \mathbf{B}_{s,q_i},
            \mathbf{Y}_{s,q_i},
            \mathbf{M}_{s,q_i},
            q_i
        \right)
    \right\}_{i=1}^{K},
    \label{eq:support_set}
\end{equation}
where each source BeamRM is paired with the query and beam map under which it was generated. Source-assisted prediction is then written as
\begin{equation}
    \widehat{\mathbf{Y}}^{\,K}_{s,q}
    =
    F_{\theta}
    \left(
        \mathbf{G}_{s},
        \mathbf{E}_{s,q},
        q,
        \mathcal{Z}^{K}_{s,q}
    \right).
    \label{eq:source_assisted}
\end{equation}

Under the strict-small protocol, every support configuration satisfies
\begin{equation}
    c_i\neq c,
    \qquad
    c_i\in\mathcal{C}_{\mathrm{small}}(c),
    \qquad i=1,\ldots,K,
    \label{eq:strict_small}
\end{equation}
where $\mathcal{C}_{\mathrm{small}}(c)$ is a predefined set of configurations with lower deployment complexity than the target configuration. Consequently, the target configuration is excluded from the support pool, no target BeamRM pixel is observed, and model parameters remain fixed during inference.

Although the source BeamRMs share the target propagation scene, they correspond to different spatial excitations. They therefore provide realized cross-configuration evidence of the common environment rather than direct observations of the target field.

Based on the above definitions, the overall BeamRM prediction task is formulated as follows.

\noindent\textbf{Problem 1:}
\begin{equation}
\begin{aligned}
\underset{\theta}{\operatorname{minimize}}\quad
&\mathbb{E}\!\left[
\mathcal{L}_{\mathrm{RM}}\!\left(
\widehat{\mathbf{Y}}^{\,K}_{s,q},
\mathbf{Y}_{s,q};\mathbf{M}_{s,q}
\right)\right] \\
\mathrm{s.t.}\quad
&\widehat{\mathbf{Y}}^{\,K}_{s,q}
=
F_{\theta}\!\left(
\mathbf{G}_{s},
\mathbf{E}_{s,q},
q,
\mathcal{Z}^{K}_{s,q}
\right), \\
&
F_{\theta}\!\left(
\mathbf{G}_{s},
\mathbf{E}_{s,q},
q,
\varnothing
\right)
=
F^{0}_{\theta}\!\left(
\mathbf{G}_{s},
\mathbf{E}_{s,q}
\right),
\end{aligned}
\label{eq:problem_formulation}
\end{equation}
where $\mathcal{L}_{\mathrm{RM}}(\cdot)$ denotes a mask-aware BeamRM reconstruction loss, with $\mathcal{Z}^{0}_{s,q}=\varnothing$ and, for $K>0$, support configurations constrained by \eqref{eq:strict_small}. Hence, $K=0$ gives query-only prediction, whereas $K>0$ permits strict-small source-assisted prediction without target BeamRM observations or inference-time parameter updates.
\section{Proposed BeamRMX Framework}
\label{sec:beamrmx}

\begin{figure*}[!t]
    \centering
    \includegraphics[width=\textwidth]{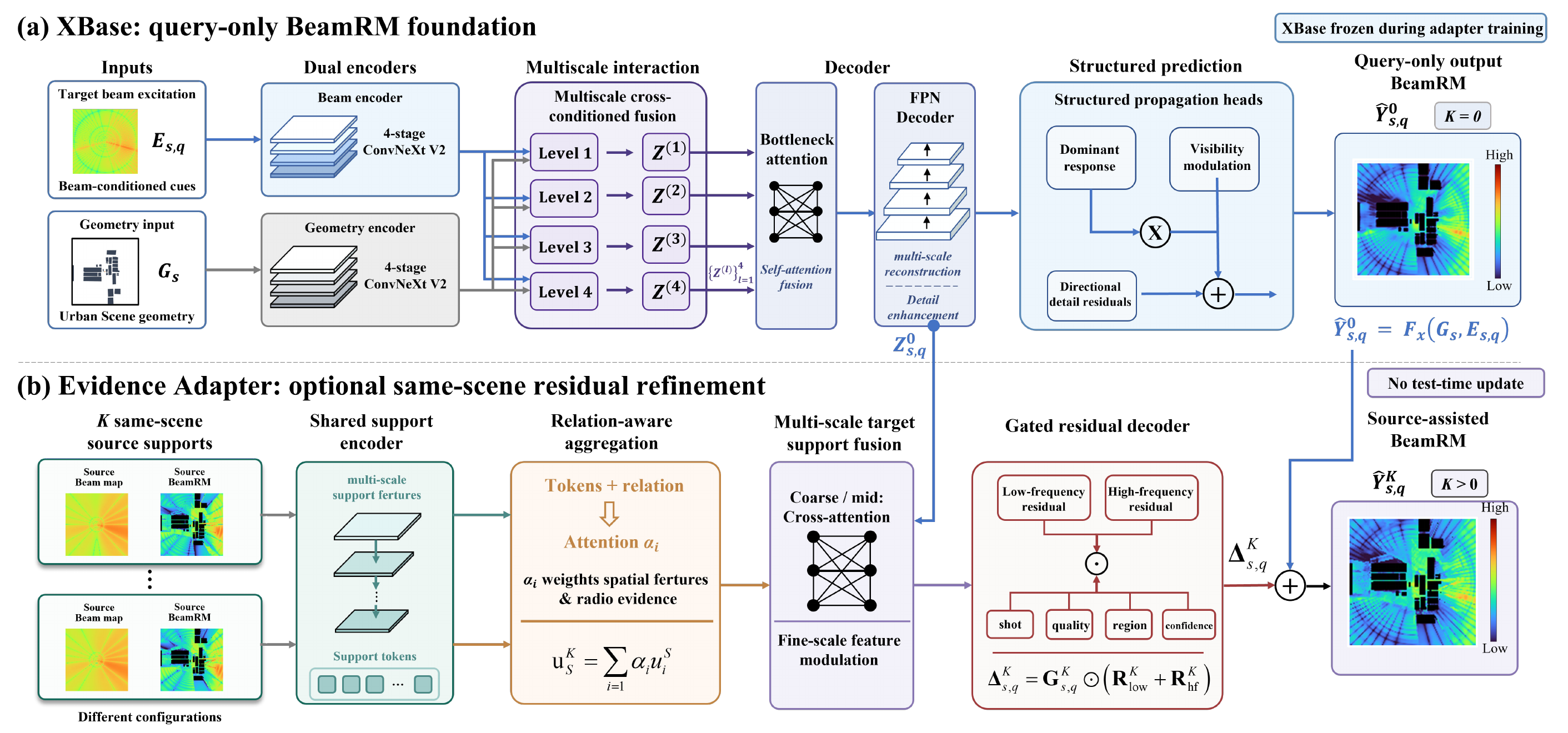}
    \caption{Architecture of BeamRMX. 
    (a) The query-only XBase learns the scene-conditioned transformation from target spatial excitation and scene geometry to the BeamRM. 
    (b) The optional Evidence Adapter exploits same-scene cross-configuration BeamRMs to produce a gated residual refinement of the frozen XBase output. 
    The adapter is bypassed for $K=0$ and activated for $K>0$, without test-time parameter updates.}
    \label{fig:beamrmx_framework}
\end{figure*}

BeamRMX instantiates the formulation in Section~III through two asymmetric components. XBase serves as the complete query-only predictor and approximates the scene-conditioned excitation-to-field transformation $\mathcal{T}_{s}$. When same-scene cross-configuration evidence is available, an optional Evidence Adapter refines the frozen XBase prediction through a gated residual correction.

Throughout this section, power-valued beam maps and BeamRMs are represented on a common affine normalized scale in $[0,1]$, and $\Pi_Y(\cdot)$ denotes element-wise clipping to this range. Because the same affine mapping is used, normalized power differences remain proportional to their dB-domain counterparts.

\subsection{Framework Overview}

Given the scene representation $\mathbf{G}_{s}$ and target spatial excitation $\mathbf{E}_{s,q}$, XBase produces
\begin{equation}
    \left(
        \widehat{\mathbf{Y}}^{\,0}_{s,q},
        \mathbf{Z}^{0}_{s,q}
    \right)
    =
    F_X
    \left(
        \mathbf{G}_{s},
        \mathbf{E}_{s,q};
        \theta_X
    \right),
    \label{eq:xbase_overview}
\end{equation}
where $\widehat{\mathbf{Y}}^{\,0}_{s,q}$ is the complete query-only BeamRM and $\mathbf{Z}^{0}_{s,q}$ is the target-side representation exposed to the Evidence Adapter.

For $K>0$, the adapter estimates a residual correction from the frozen XBase anchor and the support set $\mathcal{Z}^{K}_{s,q}$:
\begin{equation}
\mathbf{\Delta}^{K}_{s,q}
=
A\!\left(
    \widehat{\mathbf{Y}}^{\,0}_{s,q},
    \mathbf{Z}^{0}_{s,q},
    \mathbf{G}_{s},
    q,
    \mathcal{Z}^{K}_{s,q};
    \theta_A
\right),
    \label{eq:adapter_overview}
\end{equation}
\begin{equation}
    \widehat{\mathbf{Y}}^{\,K}_{s,q}
    =
    \Pi_Y\!\left(
        \widehat{\mathbf{Y}}^{\,0}_{s,q}
        +
        \mathbf{\Delta}^{K}_{s,q}
    \right),
    \qquad
   \\ 
   \mathbf{\Delta}^{0}_{s,q}=\mathbf{0}.
    \label{eq:beamrmx_output}
\end{equation}
Thus, XBase remains a standalone predictor rather than a preliminary stage requiring subsequent correction. For $K=0$, the complete support branch is structurally bypassed and BeamRMX returns $\widehat{\mathbf{Y}}^{\,0}_{s,q}$ exactly, as illustrated in Fig.~\ref{fig:beamrmx_framework}.

\subsection{XBase: Scene-Conditioned Excitation-to-Field Learning}
\label{subsec:xbase}

As shown in Fig.~\ref{fig:beamrmx_framework}(a), XBase preserves the distinct roles of spatial excitation and propagation geometry before explicitly learning their interaction. It therefore processes $\mathbf{E}_{s,q}$ and $\mathbf{G}_s$ through separate four-stage ConvNeXt V2 encoders \cite{woo2023convnextv2}:
\begin{equation}
\begin{aligned}
\left\{\mathbf{U}_{e}^{(\ell)}\right\}_{\ell=1}^{L}
&=
E_{e}\!\left(\mathbf{E}_{s,q}\right),\\
\left\{\mathbf{U}_{g}^{(\ell)}\right\}_{\ell=1}^{L}
&=
E_{g}\!\left(\mathbf{G}_{s}\right).
\end{aligned}
\label{eq:xbase_encoders}
\end{equation}
Here, $\mathbf{E}_{s,q}$ is the spatial representation of the target radiation state. Its configuration dependence is encoded through the target beam map contained in $\mathbf{E}_{s,q}$; XBase does not use an additional global query embedding. The excitation stream contains the target beam map and transmitter-relative spatial cues, whereas the geometry stream represents building and boundary-related scene structure.

In implementation, the beam-conditioned encoder receives five channels comprising the target beam map, transmitter-location map, two normalized coordinate maps, and transmitter-distance map. The geometry encoder receives 11 channels describing building height and occupancy, scene edges and boundary bands, boundary distance, horizontal and vertical height gradients, corner responses, normalized coordinates, and transmitter distance. The target BeamRM is never included in either input stream.

At each resolution level, the two streams interact through cross-conditioned fusion:
\begin{equation}
\begin{aligned}
\left[
\boldsymbol{\Gamma}_{e}^{(\ell)},
\boldsymbol{\Gamma}_{g}^{(\ell)}
\right]
&=
\sigma\!\left(
H_{\Gamma}^{(\ell)}\!\left(
\operatorname{Concat}\!\left(
\mathbf{U}_{e}^{(\ell)},
\mathbf{U}_{g}^{(\ell)}
\right)\right)\right),\\
\mathbf{Z}^{(\ell)}
&=
H_{F}^{(\ell)}\!\left(
\operatorname{Concat}\!\left(
\boldsymbol{\Gamma}_{e}^{(\ell)}\odot\mathbf{U}_{e}^{(\ell)},
\boldsymbol{\Gamma}_{g}^{(\ell)}\odot\mathbf{U}_{g}^{(\ell)}
\right)\right).
\end{aligned}
\label{eq:cross_conditioned_fusion}
\end{equation}
The learned masks condition each stream on the other before fusion, preserving their distinct information roles while modeling excitation--geometry interaction across multiple spatial scales. This contrasts with input-level concatenation, where the two information types are merged before their respective structures are encoded.

A feature pyramid network (FPN)-style decoder \cite{lin2017fpn} yields
\begin{equation}
    \mathbf{Z}^{0}_{s,q}
    =
    D_X\!\left(
    \left\{\mathbf{Z}^{(\ell)}\right\}_{\ell=1}^{L}
    \right).
    \label{eq:xbase_decoder}
\end{equation}

The decoded representation is mapped to a dominant response and
a scene-dependent modulation map:
\begin{equation}
\begin{aligned}
    \mathbf{Y}^{\mathrm{base}}_{s,q}
    &=
    H_{\mathrm{base}}
    \!\left(\mathbf{Z}^{0}_{s,q},\mathbf{E}_{s,q}\right),\\
    \mathbf{V}_{s,q}
    &=
    \sigma\!\left[
    H_{\mathrm{vis}}
    \!\left(\mathbf{Z}^{0}_{s,q},\mathbf{G}_{s}\right)
    \right].
\end{aligned}
\label{eq:xbase_base}
\end{equation}

Two residual heads further capture excitation--geometry and
boundary-sensitive corrections:
\begin{equation}
\begin{aligned}
    \mathbf{R}^{\mathrm{dir}}_{s,q}
    &=
    H_{\mathrm{dir}}
    \!\left(\mathbf{Z}^{0}_{s,q},
    \mathbf{E}_{s,q},\mathbf{G}_{s}\right),\\
    \mathbf{R}^{\mathrm{bd}}_{s,q}
    &=
    H_{\mathrm{bd}}
    \!\left(\mathbf{Z}^{0}_{s,q},\mathbf{G}_{s}\right).
\end{aligned}
\label{eq:xbase_residuals}
\end{equation}

The final query-only prediction is then
\begin{equation}
    \widehat{\mathbf{Y}}^{0}_{s,q}
    =
    \Pi_Y\!\left(
    \mathbf{V}_{s,q}\odot\mathbf{Y}^{\mathrm{base}}_{s,q}
    +
    \mathbf{R}^{\mathrm{dir}}_{s,q}
    +
    \mathbf{R}^{\mathrm{bd}}_{s,q}
    \right).
    \label{eq:xbase_output}
\end{equation}

Here, $\mathbf{V}_{s,q}$ softly modulates the dominant
beam-conditioned response according to scene structure.
The directional and boundary residual heads capture local
excitation--geometry deviations and boundary-sensitive details,
respectively. These components provide propagation-related
inductive biases rather than explicit decompositions of individual
reflection, diffraction, or other propagation mechanisms.

\subsection{Evidence Adapter: Cross-Configuration Propagation Evidence}
\label{subsec:evidence_adapter}

As shown in Fig.~\ref{fig:beamrmx_framework}(b), the Evidence Adapter is activated only when same-scene source evidence is available. For $K>0$, each support BeamRM is paired with its source excitation
$\mathbf{E}_{s,q_i}
=
\phi_{\mathrm{exc}}
\left(
\mathbf{B}_{s,q_i};
\Omega_s
\right)$.
To expose how the realized source response differs from the corresponding beam map, we construct
\begin{equation}
    \mathbf{D}^{\mathrm{bp}}_{i}
    =
    \mathbf{M}_{s,q_i}
    \odot
    \left(
        \mathbf{Y}_{s,q_i}
        -
        \mathbf{B}_{s,q_i}
    \right).
    \label{eq:support_discrepancy}
\end{equation}
This response-to-prior discrepancy is used only as a learned evidence cue and is not interpreted as an isolated physical propagation component.

The support input and its encoded representation are
\begin{equation}
\begin{aligned}
\mathbf{X}^{S}_{i}
&=
\phi_{S}\!\left(
\mathbf{E}_{s,q_i},
\mathbf{Y}_{s,q_i},
\mathbf{M}_{s,q_i},
\mathbf{D}^{\mathrm{bp}}_{i},
\mathbf{G}_{s}
\right),\\
E_S\!\left(\mathbf{X}^{S}_{i}\right)
&=
\left(
\left\{
\mathbf{U}^{S,(\ell)}_{i}
\right\}_{\ell=1}^{L_S},
\mathbf{u}^{S}_{i}
\right),
\end{aligned}
\label{eq:support_encoding}
\end{equation}
where $\phi_S(\cdot)$ also incorporates detail-, gradient-, and boundary-sensitive cues derived from the available source maps.

Different source configurations need not be equally informative. We therefore employ attention-based relation weighting \cite{vaswani2017attention}. Let $\mathbf{t}^{0}_{s,q}$ be a target token pooled from $\mathbf{Z}^{0}_{s,q}$, and let $\mathbf{r}(q,q_i)$ encode the relation between the target and source queries. Relation-aware aggregation is performed as
\begin{equation}
\begin{aligned}
\beta_i
&=
\psi\!\left(
\mathbf{t}^{0}_{s,q},
\mathbf{u}^{S}_{i},
\mathbf{r}(q,q_i)
\right),\\
\alpha_i
&=
\operatorname{softmax}_{i}\!\left(
\{\beta_j\}_{j=1}^{K}
\right),\\
\mathbf{u}^{K}_{S}
&=
\sum_{i=1}^{K}
\alpha_i\mathbf{u}^{S}_{i},\\
\mathbf{U}^{(\ell),K}_{S}
&=
\sum_{i=1}^{K}
\alpha_i\mathbf{U}^{S,(\ell)}_{i}.
\end{aligned}
\label{eq:support_aggregation}
\end{equation}
The relation descriptor includes frequency, antenna-configuration, and directional-state differences. This relation-aware weighting operates only on the already selected support items and is distinct from support selection itself. Candidate supports are first selected from the admissible strict-small pool using the query-aware utility criterion described in Section~V-C, after which \eqref{eq:support_aggregation} performs learned aggregation over the selected set.

For compact notation, let
\[
    \mathcal{U}^{K}_{S}
    =
    \left\{
        \mathbf{U}^{(\ell),K}_{S}
    \right\}_{\ell=1}^{L_S}
\]
denote the aggregated multiscale support features. The aggregated evidence is then fused with the frozen XBase representation as
\begin{equation}
\begin{aligned}
\mathbf{H}^{K}_{s,q}
&=
F_A\!\left(
\widehat{\mathbf{Y}}^{\,0}_{s,q},
\mathbf{Z}^{0}_{s,q},
\mathcal{U}^{K}_{S},
\mathbf{u}^{K}_{S}
\right),
\\
\mathbf{P}^{K}_{s,q}
&=
H_P\!\left(
\mathbf{H}^{K}_{s,q},
\widehat{\mathbf{Y}}^{\,0}_{s,q}
\right),
\\
\mathbf{C}^{K}_{s,q}
&=
\sigma\!\left[
H_C\!\left(
\mathbf{H}^{K}_{s,q},
\widehat{\mathbf{Y}}^{\,0}_{s,q}
\right)
\right],
\\
\left[
\mathbf{R}^{K}_{\mathrm{low}},
\mathbf{R}^{K}_{\mathrm{hf}}
\right]
&=
H_R\!\left(
\mathbf{H}^{K}_{s,q},
\mathbf{P}^{K}_{s,q}
\right).
\end{aligned}
\label{eq:adapter_heads}
\end{equation}
Here, $\mathbf{P}^{K}_{s,q}$ is a support-derived target estimate used internally by the adapter, while the two residual proposals capture broad calibration and localized detail corrections.

To prevent indiscriminate transfer across different excitations, the correction is controlled by shot, support-quality, region, and confidence factors:
\begin{equation}
\begin{aligned}
\gamma_K
&= g_K(K),
\\
\eta^{K}_{\mathrm{qual}}
&=
\sigma\!\left[
h_{\mathrm{qual}}\!\left(
\mathbf{t}^{0}_{s,q},
\mathbf{u}^{K}_{S}
\right)
\right],
\\
\mathbf{G}^{K}_{\mathrm{reg}}
&=
\sigma\!\left[
H_{\mathrm{reg}}\!\left(
\mathbf{H}^{K}_{s,q}
\right)
\right],
\\
\mathbf{G}^{K}_{s,q}
&=
\gamma_K\,
\eta^{K}_{\mathrm{qual}}\,
\mathbf{G}^{K}_{\mathrm{reg}}
\odot
\frac{
1+\mathbf{C}^{K}_{s,q}
}{2}.
\end{aligned}
\label{eq:evidence_gate}
\end{equation}
The final correction is
\begin{equation}
\begin{aligned}
\boldsymbol{\Delta}^{K}_{s,q}
&=
\mathbf{G}^{K}_{s,q}
\odot
\left(
\mathbf{R}^{K}_{\mathrm{low}}
+
\mathbf{R}^{K}_{\mathrm{hf}}
\right),
\\[2pt]
\widehat{\mathbf{Y}}^{\,K}_{s,q}
&=
\Pi_Y\!\left(
\widehat{\mathbf{Y}}^{\,0}_{s,q}
+
\boldsymbol{\Delta}^{K}_{s,q}
\right).
\end{aligned}
\label{eq:evidence_correction}
\end{equation}
The gating mechanism restricts cross-configuration correction when the available evidence is weak, spatially irrelevant, or inconsistent with the target query.

\subsection{Two-Stage Learning and Inference}
\label{subsec:learning_inference}

BeamRMX is trained in two stages. XBase is first optimized independently as the query-only predictor:
\begin{equation}
    \mathcal{L}_{X}
    =
    \mathcal{L}^{X}_{\mathrm{valid}}
    +
    \lambda_{\mathrm{aux}}
    \mathcal{L}^{X}_{\mathrm{aux}}
    +
    \lambda_{\mathrm{str}}
    \mathcal{L}^{X}_{\mathrm{str}},
    \label{eq:xbase_loss}
\end{equation}
where the terms supervise valid-region reconstruction, auxiliary structured outputs, and region/detail-sensitive consistency. These objectives require no path-level physical labels.

After this stage, $\theta_X$ is frozen and only the Evidence Adapter is trained. Episodic training samples a target query and a strict-small support set with varying $K$, using
\begin{equation}
\begin{split}
    \mathcal{L}_{A}
    ={}&
    \mathcal{L}^{A}_{\mathrm{valid}}
    +
    \lambda_{\mathrm{det}}
    \mathcal{L}^{A}_{\mathrm{det}}
    +
    \lambda_{\mathrm{prior}}
    \mathcal{L}_{\mathrm{prior}}
    +
    \lambda_{\mathrm{gain}}
    \mathcal{L}_{\mathrm{gain}}
    \\
    &+
    \lambda_{\mathrm{sp}}
    \mathcal{L}_{\mathrm{sp}}
    +
    \lambda_{\mathrm{nh}}
    \mathcal{L}_{\mathrm{nh}}
    +
    \lambda_{\mathrm{kd}}
    \mathcal{L}_{\mathrm{kd}} .
\end{split}
\label{eq:adapter_loss}
\end{equation}
The first two terms supervise final reconstruction and spatial detail; $\mathcal{L}_{\mathrm{prior}}$ and $\mathcal{L}_{\mathrm{gain}}$ train useful evidence-conditioned estimates and corrections; $\mathcal{L}_{\mathrm{sp}}$ limits unnecessarily broad adaptation; and $\mathcal{L}_{\mathrm{kd}}$ transfers more reliable richer-support behavior to low-shot settings.

A margin-based no-harm term explicitly discourages substantial degradation relative to the frozen anchor. Let
\[
    \mathcal{V}_{s,q}
    =
    \left\{
        p\in\Omega_s:
        \mathbf{M}_{s,q}(p)=1
    \right\}
\]
denote the valid-pixel set, and define the pointwise error as
$\varepsilon^{K}_{s,q}(p)
=
|\widehat{\mathbf{Y}}^{\,K}_{s,q}(p)-\mathbf{Y}_{s,q}(p)|$,
with $\varepsilon^{0}_{s,q}(p)$ defined analogously. The no-harm penalty is
\begin{equation}
    \mathcal{L}_{\mathrm{nh}}
    =
    \frac{1}{|\mathcal{V}_{s,q}|}
    \sum_{p\in\mathcal{V}_{s,q}}
    \left[
        \varepsilon^{K}_{s,q}(p)
        -
        \varepsilon^{0}_{s,q}(p)
        -
        \epsilon_{\mathrm{nh}}
    \right]_{+}.
    \label{eq:noharm_loss}
\end{equation}
Here, $[x]_{+}=\max(x,0)$. The term penalizes harmful corrections beyond the tolerance $\epsilon_{\mathrm{nh}}$ without requiring every corrected pixel to improve over XBase.

During inference, both $\theta_X$ and $\theta_A$ remain fixed. For $K=0$, the support branch is bypassed and the query-only prediction is returned exactly. For $K>0$, XBase is evaluated once and the Evidence Adapter produces the gated correction in a single forward pass. No gradient update, target-map optimization, or test-time fine-tuning is performed.
\section{Experimental Results and Discussion}
\label{sec:experiments}

This section evaluates BeamRMX from three complementary perspectives. First, we examine query-only prediction under matched-domain, scene zero-shot (Scene-ZS), and configuration zero-shot (Config-ZS) settings. Second, we evaluate whether a few same-scene BeamRMs from lower-complexity configurations can improve prediction of an unobserved target BeamRM without target observations or test-time parameter updates. Third, we analyze which evidence mechanisms are responsible for the source-assisted gain and whether the resulting BeamRM improvements translate into more reliable intra-sector beam refinement.

\begin{table*}[!t]
    \centering
    \caption{Query-only BeamRM prediction under the matched-domain, Scene-ZS, and Config-ZS protocols.}
    \label{tab:query_only_results}

    \footnotesize
    \renewcommand{\arraystretch}{1.10}

    \begin{tabular*}{\textwidth}{
        @{\extracolsep{\fill}}
        llcccccc
        @{}
    }
        \toprule
        \textbf{Protocol} &
        \textbf{Metric} &
        \textbf{RadioWNet} &
        \textbf{BeamCKM} &
        \textbf{RadioDiff} &
        \textbf{BeamCKMDiff} &
        \textbf{BeamRMX} &
        \shortstack{\textbf{Gain over}\\[-1pt]\textbf{Best Baseline}} \\
        \midrule

        \multirow{4}{*}{\shortstack[l]{Matched-\\domain}}
        & MAE (dB) $\downarrow$
        & 6.1612
        & \second{4.8713}
        & 5.0010
        & 7.1315
        & \best{2.0991}
        & \gainlow{56.91\%} \\

        & RMSE (dB) $\downarrow$
        & 9.5377
        & \second{6.9161}
        & 7.0782
        & 10.4604
        & \best{4.1597}
        & \gainlow{39.85\%} \\

        & PSNR (dB) $\uparrow$
        & 30.9427
        & \second{33.0515}
        & 32.7574
        & 29.7024
        & \best{37.1613}
        & \gainhigh{4.1098 dB} \\

        & SSIM $\uparrow$
        & 0.7015
        & \second{0.8216}
        & 0.8120
        & 0.7468
        & \best{0.9073}
        & \gainhigh{0.0857} \\
        \midrule

        \multirow{4}{*}{Scene-ZS}
        & MAE (dB) $\downarrow$
        & 6.9671
        & 6.7375
        & \second{6.7331}
        & 7.8319
        & \best{4.9749}
        & \gainlow{26.11\%} \\

        & RMSE (dB) $\downarrow$
        & 10.4689
        & \second{9.7920}
        & 9.9510
        & 11.7170
        & \best{9.0415}
        & \gainlow{7.66\%} \\

        & PSNR (dB) $\uparrow$
        & 29.9977
        & 30.2605
        & \second{30.2803}
        & 28.8026
        & \best{31.5072}
        & \gainhigh{1.2269 dB} \\

        & SSIM $\uparrow$
        & 0.6622
        & 0.7592
        & \second{0.7679}
        & 0.7200
        & \best{0.8106}
        & \gainhigh{0.0427} \\
        \midrule

        \multirow{4}{*}{Config-ZS}
        & MAE (dB) $\downarrow$
        & 6.1466
        & 5.4694
        & \second{5.3107}
        & 8.6834
        & \best{2.7748}
        & \gainlow{47.75\%} \\

        & RMSE (dB) $\downarrow$
        & 8.8138
        & 7.5310
        & \second{7.3779}
        & 11.9391
        & \best{5.0562}
        & \gainlow{31.47\%} \\

        & PSNR (dB) $\uparrow$
        & 31.3766
        & 32.2486
        & \second{32.4379}
        & 28.4284
        & \best{35.4660}
        & \gainhigh{3.0281 dB} \\

        & SSIM $\uparrow$
        & 0.6976
        & 0.8006
        & \second{0.8245}
        & 0.7179
        & \best{0.8695}
        & \gainhigh{0.0450} \\
        \bottomrule
    \end{tabular*}

    \vspace{2pt}
\end{table*}
\subsection{Experimental Setup}
\label{subsec:experimental_setup}

Experiments are conducted on the multi-configuration XL-MIMO radiomap dataset introduced in \cite{li2026u6gxlmimoradiomapprediction}. The dataset contains 78,400 ray-tracing-based BeamRMs from 800 urban scenes, covering five carrier frequencies from 1.8 to 6.7~GHz, multiple array geometries and sizes, and directional arrays with up to 1024 antenna elements. The combinations of carrier frequency, array configuration, and beam steering state yield 98 concrete frequency--array--beam query conditions per scene. Each sample consists of a scene representation, a configuration-dependent beam map, and the corresponding BeamRM.

All scene representations, beam maps, and target BeamRMs are processed at a common spatial resolution of $128\times128$. Valid free-space received powers within $(-300,0)$~dB are linearly mapped to $[0,1]$, while building and invalid regions are excluded using the corresponding validity masks. Mean absolute error (MAE) and root mean square error (RMSE) are computed over valid free-space pixels after conversion back to the dB domain, whereas peak signal-to-noise ratio (PSNR) and structural similarity index (SSIM) are evaluated on the normalized BeamRMs with unit data range and the same valid-region treatment. Unless otherwise stated, all compared methods use identical data splits, spatial resolution, normalization, validity masks, and metric implementations.

All experiments are implemented in PyTorch and conducted on NVIDIA A40 graphics processing units (GPUs). XBase employs four-stage ConvNeXt V2 encoders and an FPN-style decoder and is optimized using AdamW with mixed-precision training for 50 epochs, with checkpoints selected according to validation MAE. The Evidence Adapter is subsequently trained for 16 epochs while XBase remains frozen, using a learning rate of $5\times10^{-5}$ and a weight decay of $2\times10^{-2}$. During episodic adapter training, $K\in\{0,1,2,4,8\}$ is sampled with probabilities $\{0.05,0.45,0.35,0.10,0.05\}$, emphasizing the practically relevant low-shot regime. No model performs target-side parameter updates during inference. Complete training configurations and evaluation scripts are provided in the public code repository.

Model scale and inference efficiency:
XBase contains 89.3M parameters, comparable in scale to the 105.6M–113.0M adapted BeamCKM-based predictors used in our experiments. The optional Evidence Adapter introduces 42.3M trainable parameters, giving 131.6M parameters for the complete source-assisted model while leaving XBase frozen. Under our NVIDIA A40 evaluation implementation, query-only XBase requires approximately 50.2~ms per query, and source-assisted inference requires approximately 60--62~ms for $K\in\{1,2,4\}$. Although the evaluated diffusion-based predictors contain fewer parameters, their iterative sampling pipelines require approximately 115--121~ms per query. Thus, the proposed framework retains single-pass inference and introduces only a modest latency increase when cross-configuration evidence is available.

\subsection{Query-Only Multi-Configuration Prediction}
\label{subsec:query_only_results}

We first evaluate the query-only prediction capability of BeamRMX before introducing any same-scene radio evidence. Each method is provided with the scene representation, transmitter information, and the spatial beam map associated with the target query, but has no access to same-scene source BeamRMs, sparse target measurements, partial target maps, or target labels. This setting isolates the ability of each model to learn the scene-conditioned transformation from a target spatial-excitation prior to the corresponding received power field.

The comparison includes RadioWNet + Beam Map, BeamCKM, RadioDiff, and BeamCKMDiff. RadioWNet follows the beam-conditioned W-Net benchmark established in the multi-configuration dataset study \cite{li2026u6gxlmimoradiomapprediction}. The adapted BeamCKM baseline uses a deterministic encoder--decoder predictor with beam-conditioned inputs \cite{wang2025beamckm}, whereas RadioDiff represents a diffusion-based radio map predictor \cite{wang2024radiodiff}. BeamCKMDiff combines continuous beam-aware conditioning with a diffusion-transformer architecture \cite{zhao2026beamckmdiff,peebles2023dit}. For a representation-controlled comparison, all methods are provided with the same target beam map as the spatial beam representation. BeamCKM and BeamCKMDiff are therefore adapted from their original beam descriptors to the same beam map input while retaining their principal prediction architectures. This setting isolates the learning architecture from differences in beam representation.

Three complementary query-only protocols are considered:

\begin{itemize}
    \item Matched-domain:
    The 78,400 scene--query samples are randomly divided into 54,880 training, 7,840 validation, and 15,680 test samples. Scene and radio-query distributions may overlap across the subsets, while individual samples remain disjoint. This setting serves as a seen-domain reference.

    \item Scene-ZS:
    The 800 scenes are divided into 560 training, 80 validation, and 160 test scenes, with all 98 frequency--array--beam query conditions retained for each scene. The resulting 15,680 test samples are therefore generated exclusively from propagation scenes absent from model training.

    \item Config-ZS:
    Complete radio configurations are separated across training, validation, and testing while all 800 scenes are retained. A 6.7-GHz, 64-element, 8-beam configuration is used exclusively for validation, whereas a distinct 6.7-GHz, 256-element, 16-beam configuration is reserved for testing. All remaining configurations are used for training, resulting in 59,200 training, 6,400 validation, and 12,800 test samples. 
\end{itemize}

Scene-ZS and Config-ZS probe complementary generalization dimensions and are therefore not intended as directly comparable measures of difficulty. The former withholds propagation scenes, whereas the latter withholds an entire radio configuration. In particular, neither the Config-ZS test configuration nor any BeamRM associated with its directional states is used for model training or checkpoint selection.

\begin{figure*}[!t]
    \centering

    \setlength{\tabcolsep}{1.0pt}
    \renewcommand{\arraystretch}{1.0}

    \begin{tabular}{
        @{}
        >{\centering\arraybackslash}m{\qualrowlabelwidth}
        *{6}{>{\centering\arraybackslash}m{\qualpanelwidth}}
        @{}
    }

        &
        \footnotesize\bfseries Ground Truth
        &
        \footnotesize\bfseries RadioWNet
        &
        \footnotesize\bfseries BeamCKM
        &
        \footnotesize\bfseries RadioDiff
        &
        \footnotesize\bfseries BeamCKMDiff
        &
        \shortstack{
            \footnotesize\bfseries BeamRMX\\[-1pt]
            \footnotesize\bfseries (Our method)
        }
        \\[2pt]

        \footnotesize\bfseries \shortstack{Matched\\domain}
        &
        \includegraphics[width=\qualpanelwidth]{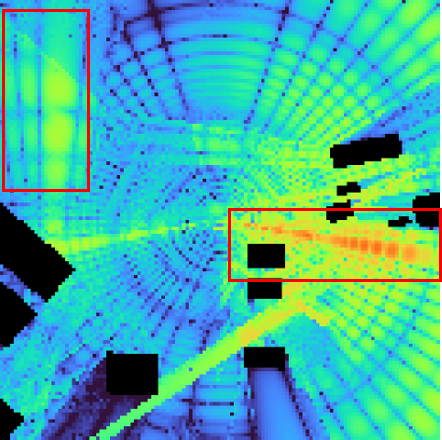}
        &
        \includegraphics[width=\qualpanelwidth]{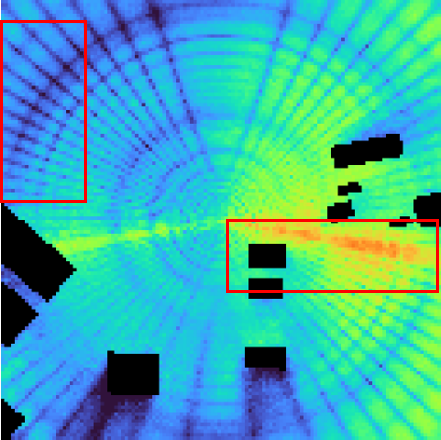}
        &
        \includegraphics[width=\qualpanelwidth]{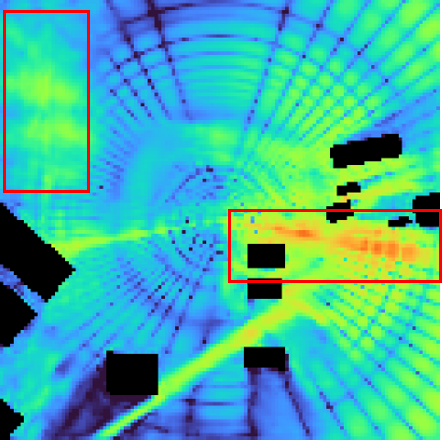}
        &
        \includegraphics[width=\qualpanelwidth]{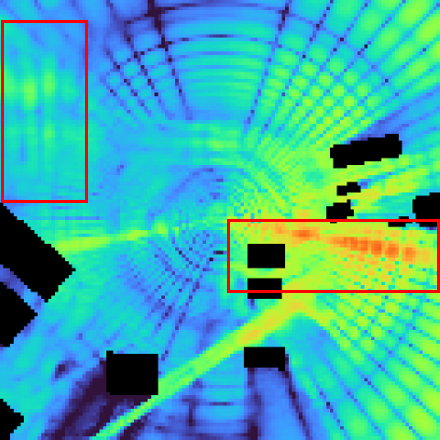}
        &
        \includegraphics[width=\qualpanelwidth]{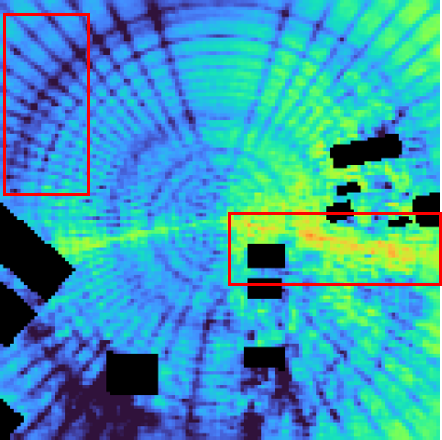}
        &
        \includegraphics[width=\qualpanelwidth]{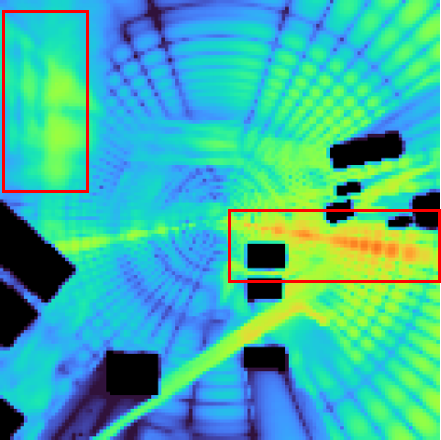}
        \\

        \noalign{\vskip 3pt}

        \footnotesize\bfseries Scene-ZS
        &
        \includegraphics[width=\qualpanelwidth]{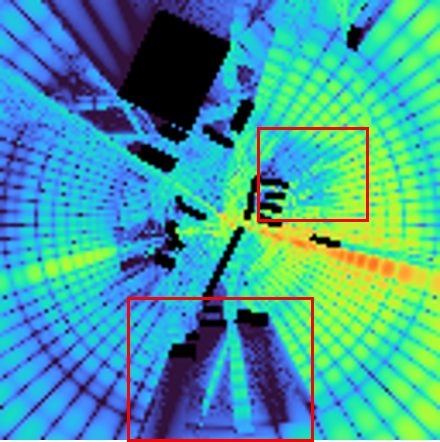}
        &
        \includegraphics[width=\qualpanelwidth]{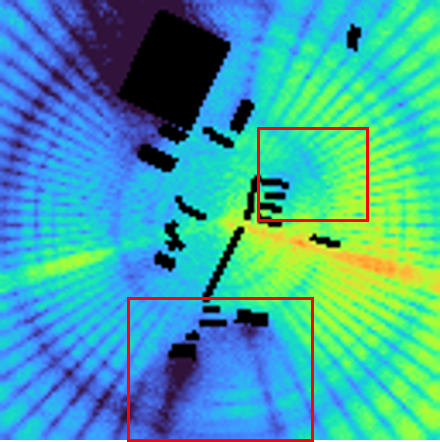}
        &
        \includegraphics[width=\qualpanelwidth]{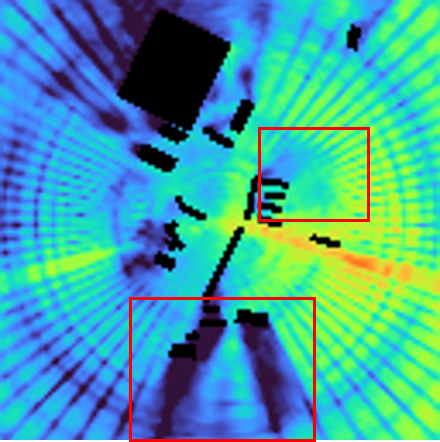}
        &
        \includegraphics[width=\qualpanelwidth]{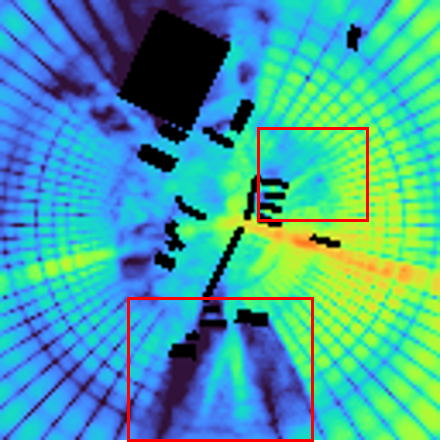}
        &
        \includegraphics[width=\qualpanelwidth]{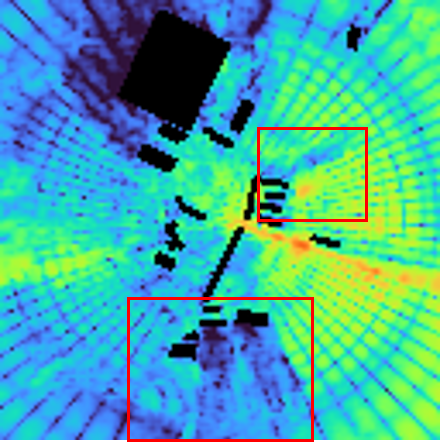}
        &
        \includegraphics[width=\qualpanelwidth]{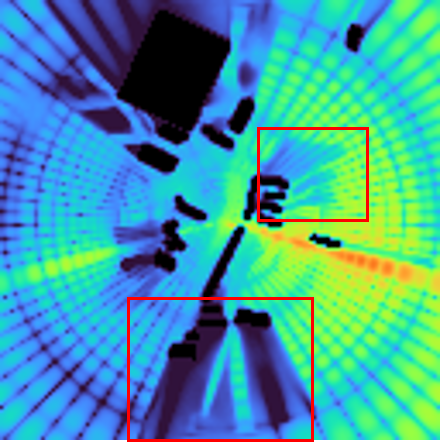}
        \\[-1pt]

        \footnotesize\bfseries Scene-ZS
        &
        \includegraphics[width=\qualpanelwidth]{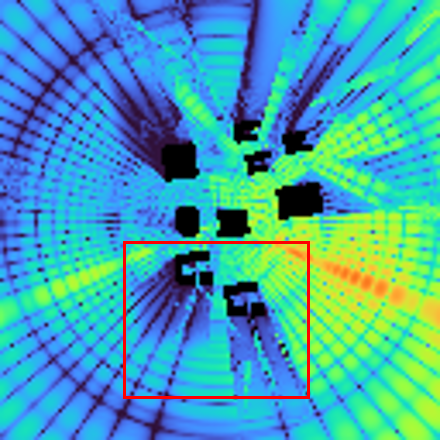}
        &
        \includegraphics[width=\qualpanelwidth]{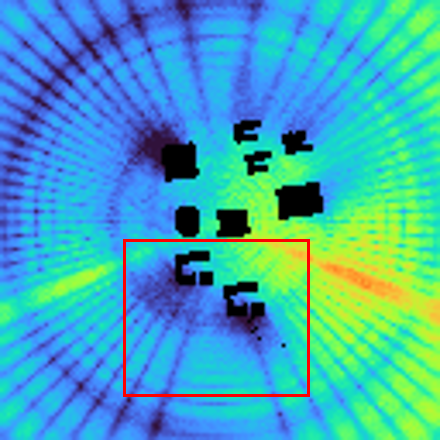}
        &
        \includegraphics[width=\qualpanelwidth]{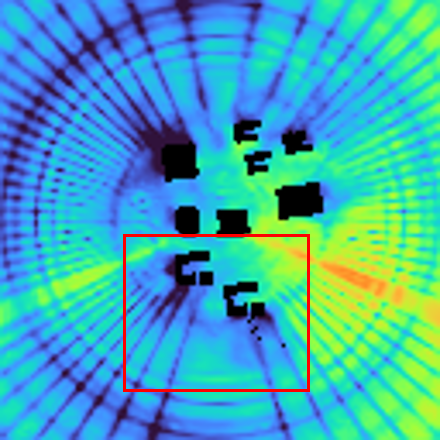}
        &
        \includegraphics[width=\qualpanelwidth]{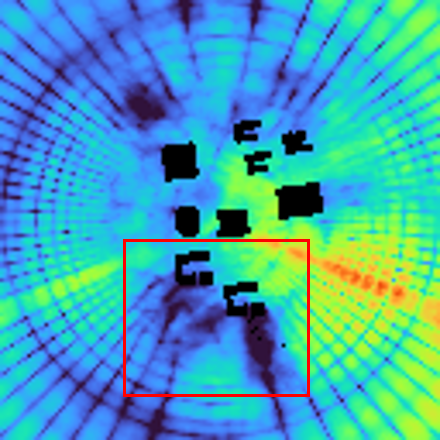}
        &
        \includegraphics[width=\qualpanelwidth]{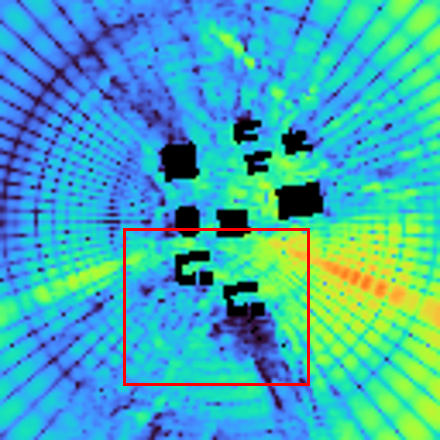}
        &
        \includegraphics[width=\qualpanelwidth]{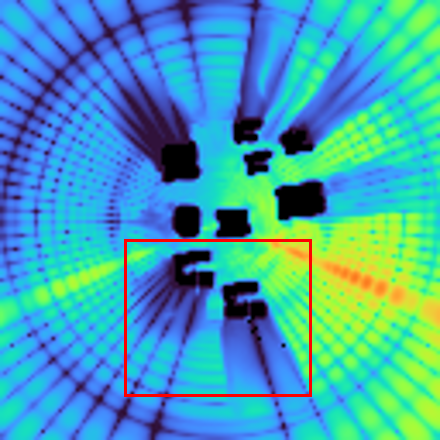}
        \\

        \noalign{\vskip 4pt}

        \footnotesize\bfseries Config-ZS
        &
        \includegraphics[width=\qualpanelwidth]{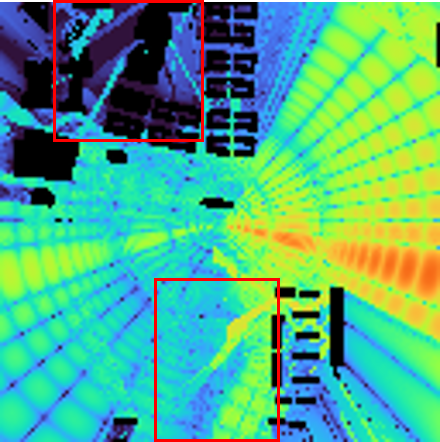}
        &
        \includegraphics[width=\qualpanelwidth]{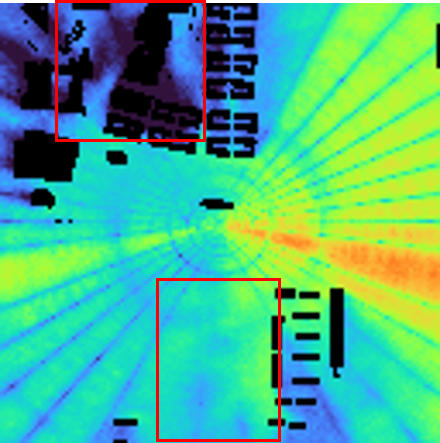}
        &
        \includegraphics[width=\qualpanelwidth]{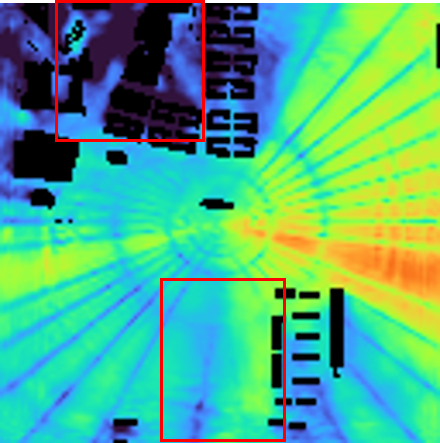}
        &
        \includegraphics[width=\qualpanelwidth]{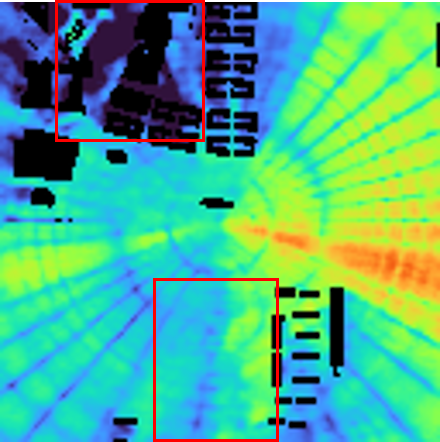}
        &
        \includegraphics[width=\qualpanelwidth]{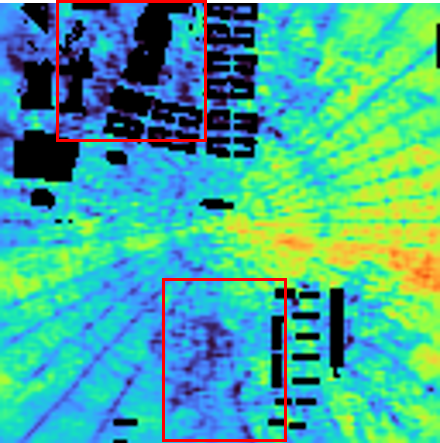}
        &
        \includegraphics[width=\qualpanelwidth]{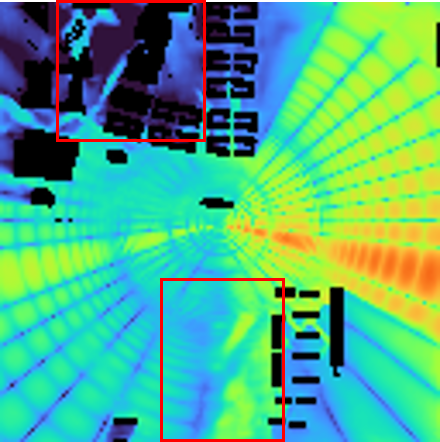}
        \\[-1pt]

        \footnotesize\bfseries Config-ZS
        &
        \includegraphics[width=\qualpanelwidth]{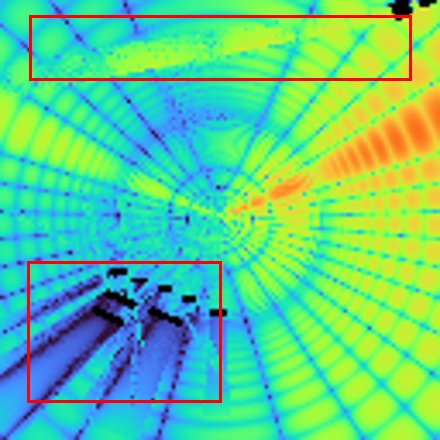}
        &
        \includegraphics[width=\qualpanelwidth]{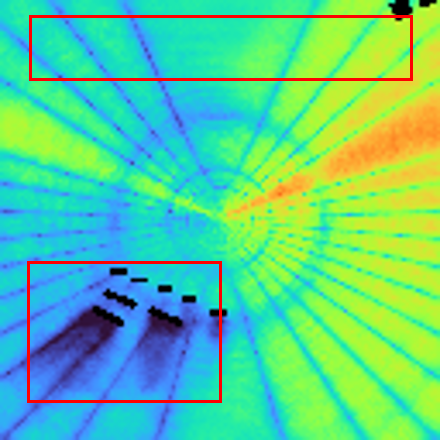}
        &
        \includegraphics[width=\qualpanelwidth]{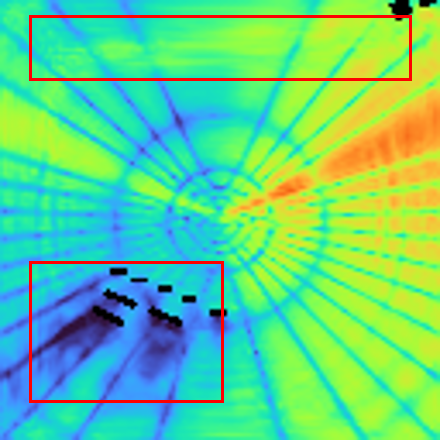}
        &
        \includegraphics[width=\qualpanelwidth]{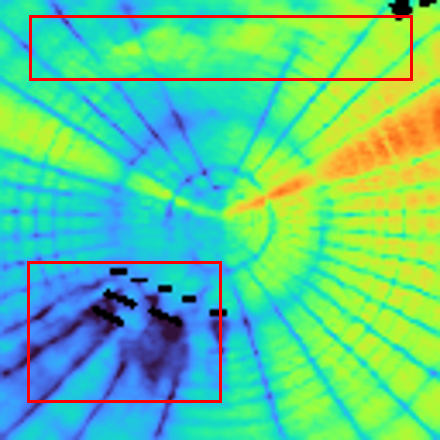}
        &
        \includegraphics[width=\qualpanelwidth]{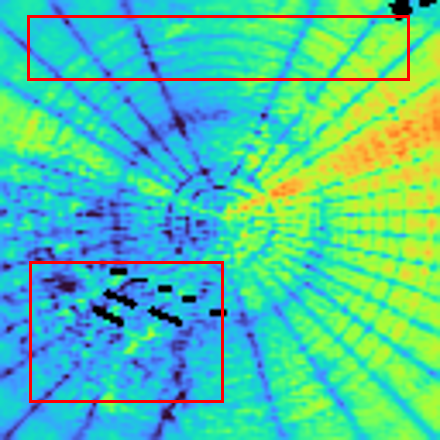}
        &
        \includegraphics[width=\qualpanelwidth]{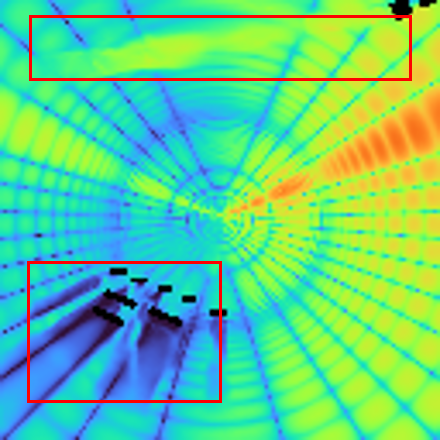}

    \end{tabular}

    \caption{Qualitative BeamRM prediction under matched-domain, Scene-ZS, and Config-ZS evaluation protocols. The figure shows one representative matched-domain example, two Scene-ZS examples, and two Config-ZS examples. All methods use the same received-power scale within each row.}
    \label{fig:qualitative_beamrm}
\end{figure*}

Table~\ref{tab:query_only_results} first shows that BeamRMX is already a strong query-only predictor in the matched-domain setting. It achieves an MAE of 2.0991~dB, reducing the best competing MAE by 56.91\%, while the RMSE is reduced by 39.85\%. The consistent advantage in PSNR and SSIM further shows that the gain is not limited to an overall power offset. Since no distribution shift is introduced in this protocol, these results establish the effectiveness of XBase as a dedicated predictor for spatial-excitation-conditioned BeamRM learning.

The Scene-ZS results test whether this learned excitation-to-field relationship transfers to completely unseen propagation environments. BeamRMX achieves an MAE of 4.9749~dB, corresponding to a 26.11\% reduction relative to the best competing result. It also obtains the best RMSE, PSNR, and SSIM among all evaluated methods. Notably, the relative RMSE reduction is more moderate at 7.66\% than the MAE improvement. This suggests that, although BeamRMX substantially improves the overall spatial prediction accuracy, difficult high-error regions remain challenging when the propagation environment itself is unseen. The result therefore supports the value of the proposed excitation--geometry interaction while also revealing residual scene-dependent uncertainty that cannot be fully resolved from geometry and the target excitation alone.

Config-ZS evaluates a different form of extrapolation: the complete 6.7-GHz, 256-element, 16-beam test configuration is absent from both model training and checkpoint selection, although the underlying scenes remain represented through other configurations. Under this configuration-disjoint protocol, BeamRMX attains an MAE of 2.7748~dB and an RMSE of 5.0562~dB, reducing the corresponding errors of the best baseline by 47.75\% and 31.47\%, respectively. It also improves PSNR from 32.4379 to 35.4660~dB and SSIM from 0.8245 to 0.8695, corresponding to gains of 3.0281~dB and 0.0450, respectively. These consistent improvements in both power-domain errors and structural metrics indicate that the Config-ZS advantage is not limited to overall received-power calibration, but also extends to preserving the beam-dependent spatial response. More importantly, the results show that the spatial-excitation formulation can transfer to a radio configuration that is entirely absent during model development, rather than relying on configuration-specific associations learned from the training samples.

Training-seed robustness:
To assess sensitivity to training stochasticity, we independently
train BeamRMX with three random seeds while keeping the Scene-ZS
and Config-ZS data partitions and all other experimental settings
fixed. The resulting Scene-ZS MAE is $4.9359 \pm 0.0344$ dB,
where the variation denotes the sample standard deviation across
independent training runs. Under Config-ZS, the corresponding MAE
is $2.8567 \pm 0.0787$ dB. In both regimes, the
run-to-run variation is substantially smaller than the performance
margin over the strongest competing method in Table I, indicating
that the observed zero-shot gains are insensitive to training
initialization and stochastic optimization.

Representative Scene-ZS and Config-ZS predictions are compared in Fig.~\ref{fig:qualitative_beamrm}. The examples complement the aggregate results by directly contrasting the reconstructed BeamRMs under identical target queries. All methods recover the dominant beam-conditioned spatial response to some extent, whereas BeamRMX more closely follows the ground-truth distribution in localized high- and low-response regions. The visual comparison is therefore consistent with the quantitative gains in Table~\ref{tab:query_only_results}, while the common received power scale prevents independent normalization from exaggerating apparent differences.

Across the three protocols, baseline performance shifts noticeably with the evaluation regime. BeamCKM is the strongest baseline in the matched-domain setting, whereas RadioDiff proves more resilient under distribution shift, taking second place across all four Config-ZS metrics. This contrast highlights a key trade-off: while deterministic designs like BeamCKM excel when training and test conditions align, diffusion-based approaches retain stronger generalization. Adapting BeamCKMDiff from vector conditioning to a spatial beam-map representation reduces its effectiveness, reflecting its architectural difficulty in handling heterogeneous multi-scene setup. Across all settings, BeamRMX consistently achieves top scores on every metric, confirming its adaptability to both unseen scenes and unseen radio configurations. Nevertheless, the performance gap between matched-domain and Scene-ZS settings points to residual scene uncertainty that geometry alone cannot resolve—providing a clear rationale for incorporating same-scene cross-configuration evidence.

\begin{table*}[!t]
    \centering
    \caption{Source-assisted BeamRM prediction under the strict-small protocol on 160 unseen scenes. }
    \label{tab:source_assisted_results}
    \footnotesize
    \setlength{\tabcolsep}{2.5pt}
    \renewcommand{\arraystretch}{1.10}
    \begin{tabular*}{\textwidth}{@{\extracolsep{\fill}}lcccccccccccc@{}}
        \toprule
        \multirow{2}{*}{\textbf{Metric}} &
        \multicolumn{4}{c}{\textbf{BeamCKM-Support}} &
        \multicolumn{4}{c}{\textbf{RadioDiff-Support}} &
        \multicolumn{4}{c}{\textbf{BeamRMX}} \\
        \cmidrule(lr){2-5}
        \cmidrule(lr){6-9}
        \cmidrule(lr){10-13}
        &
        $K=0$ & $K=1$ & $K=2$ & $K=4$ &
        $K=0$ & $K=1$ & $K=2$ & $K=4$ &
        $K=0$ & $K=1$ & $K=2$ & $K=4$ \\
        \midrule

        MAE (dB) $\downarrow$
        & 6.7336
        & \second{4.7637}
        & \second{4.7635}
        & \second{4.7635}
        & \second{6.4681}
        & 5.2391
        & 5.1547
        & 5.1850
        & \best{5.9665}
        & \best{3.9178}
        & \best{3.7237}
        & \best{3.7095} \\

        RMSE (dB) $\downarrow$
        & 9.9306
        & \second{7.0330}
        & \second{7.0328}
        & \second{7.0329}
        & \second{9.8527}
        & 7.7761
        & 7.6632
        & 7.7083
        & \best{9.5182}
        & \best{6.7461}
        & \best{6.4150}
        & \best{6.3883} \\

        PSNR (dB) $\uparrow$
        & 30.1792
        & \second{33.1113}
        & \second{33.1116}
        & \second{33.1116}
        & \second{30.4214}
        & 32.4092
        & 32.5434
        & 32.4960
        & \best{30.7559}
        & \best{33.8350}
        & \best{34.2372}
        & \best{34.2732} \\

        SSIM $\uparrow$
        & 0.7601
        & \second{0.8354}
        & \second{0.8354}
        & \second{0.8354}
        & \second{0.7764}
        & 0.8209
        & 0.8240
        & 0.8232
        & \best{0.7776}
        & \best{0.8818}
        & \best{0.8915}
        & \best{0.8924} \\

        \bottomrule
    \end{tabular*}
\end{table*}
\subsection{Strict-Small Source-Assisted Prediction in Unseen Scenes}
\label{subsec:source_assisted_results}

We next investigate whether limited same-scene radio evidence can reduce the residual uncertainty of query-only prediction in unseen propagation environments. The experiment uses the 160 Scene-ZS test scenes and focuses on the 6.7-GHz, 1024-element, 64-beam target configuration, yielding 10,240 target queries. Support BeamRMs are drawn from the same scene but from different radio configurations containing no more than 64 antenna elements. The target configuration is excluded from the support pool, and no sparse observation or partial map of the target BeamRM is available. Under this strict-small protocol, we evaluate $K\in\{0,1,2,4\}$ supports without any inference-time parameter update. Thus, the target uses a 1024-element array, whereas every admissible support configuration contains at most 64 antenna elements, with the target configuration excluded by construction.

For each target query, admissible supports are first restricted by the strict-small constraint and then ranked using a query-aware utility criterion. The utility combines source--target beam and configuration relations with source-side validity and detail cues, consistency with the frozen query-only anchor, and a redundancy penalty among selected supports. The top-$K$ candidates form the support set used by the Evidence Adapter. The selector uses only the target query, the frozen query-only prediction, and available source-side information; no target BeamRM value is used.

For each method, $K=0$ denotes its query-only anchor evaluated on the same target configuration. In BeamRMX, the Evidence Adapter is bypassed exactly at $K=0$, so the output is identical to that of the frozen XBase. The resulting $K=0$ performance differs from the Scene-ZS result in Table~\ref{tab:query_only_results}, which averages over all radio-query conditions, whereas the present experiment evaluates only the 6.7-GHz, 1024-element, 64-beam target configuration.

To determine whether the benefit of same-scene evidence is specific to BeamRMX, we construct source-assisted variants of two complementary baselines. BeamCKM is selected as the closest beam-aware reference because its M3ChanNet extension already incorporates environmental profiles and multi-beam observations as auxiliary information \cite{wang2025beamckm}, while RadioDiff represents a strong generative RM predictor and one of the most competitive zero-shot baselines in Section~V-B. Both are adapted to the same strict-small information boundary and receive exactly the same query-aware support episodes as BeamRMX. BeamCKM-Support incorporates the source BeamRMs into its deterministic pipeline, whereas RadioDiff-Support augments its diffusion condition with aggregated source observations. All methods use identical target queries, support cardinalities, support episodes, and metrics.


Table~\ref{tab:source_assisted_results} shows that even a very small amount of same-scene cross-configuration evidence provides substantial information beyond the query-only prediction. For BeamRMX, a single support reduces MAE from 5.9665 to 3.9178~dB, corresponding to a 34.3\% reduction. RMSE, PSNR, and SSIM improve simultaneously, indicating that the benefit is not limited to a global power adjustment but extends to the reconstructed spatial response. Since the target BeamRM itself remains completely unobserved, these gains show that source configurations contain scene-specific information that is transferable across different spatial excitations.

The benefit also exhibits a clear small-shot saturation trend. Increasing the support size from $K=1$ to $K=2$ further reduces the BeamRMX MAE from 3.9178 to 3.7237~dB, whereas increasing $K$ to 4 yields only a marginal additional improvement to 3.7095~dB. The other reconstruction metrics follow the same trend. Hence, most of the useful same-scene information is recovered from the first one or two informative supports, while additional configurations provide diminishing returns. This property is particularly relevant to incremental network deployment, where a small amount of legacy or lower-complexity radio information may already be available but exhaustive characterization of the new large-array configuration is undesirable.

BeamRMX also exploits the available evidence more effectively than the two support-enhanced baselines. At $K=4$, its MAE is 22.1\% lower than that of BeamCKM-Support and 28.5\% lower than that of RadioDiff-Support. More importantly, the methods respond differently as additional evidence becomes available. BeamCKM-Support essentially saturates after the first support, while RadioDiff-Support improves from $K=1$ to $K=2$ but gains no further benefit at $K=4$. BeamRMX instead retains consistent, although diminishing, improvement as support increases. This suggests that the value of source evidence depends not only on its availability, but also on whether configuration-specific response patterns can be separated from scene information that remains useful for the target query.

These results support the central design principle of the Evidence Adapter. Same-scene source BeamRMs should not be treated as direct approximations of the target field; rather, they provide configuration-conditioned observations of a shared propagation environment. Relation-aware evidence aggregation and gated residual refinement allow BeamRMX to exploit this information while preserving the frozen query-only prediction as a stable anchor. The following subsection examines which components are responsible for this gain and whether the resulting BeamRM improvements translate into more reliable beam management decisions.

\begin{figure*}[!t]
    \centering
    \includegraphics[width=\textwidth]{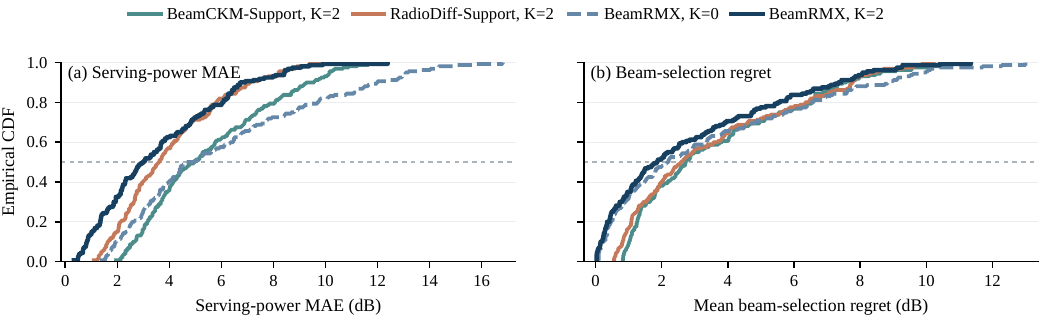}
    \caption{Scene-level empirical cumulative distribution functions (CDFs) of serving power MAE and mean beam-selection regret over 160 unseen scenes.}
    \label{fig:communication_ecdf}
\end{figure*}

\subsection{Evidence Mechanisms and Intra-Sector Beam Refinement Utility}
\label{subsec:mechanism_utility}

We evaluate the Evidence Adapter using controlled mechanism diagnostics and then examine whether the reconstruction gains improve intra-sector beam refinement.

Evidence mechanism analysis:
The 40-scene subset is sampled uniformly at random once from the 160 unseen test scenes and fixed for all mechanism diagnostics, without reference to model outputs or scene-level performance. All $K=2$ variants use the same trained BeamRMX checkpoint and the same strict-small support pool. The diagnostics respectively remove the source BeamRM content, disable the spatial region gate, or replace the query-aware utility selection with uniformly random admissible supports. High-frequency (HF) MAE is a detail-sensitive dB-domain reconstruction metric evaluated over valid free-space regions using the high-frequency criterion implemented in our evaluation code, while the no-harm violation rate denotes the fraction of valid pixels for which source-assisted prediction degrades the frozen XBase anchor beyond the prescribed tolerance. These fixed checkpoint experiments are intended to probe mechanism trends on a controlled unseen-scene subset rather than to provide a full-test ranking of independently retrained model variants.
\begin{table}[!t]
    \centering
    \caption{Fixed checkpoint mechanism diagnostics on 40 unseen scenes.}
    \label{tab:mechanism_analysis}
    \footnotesize
    \setlength{\tabcolsep}{4.2pt}
    \renewcommand{\arraystretch}{1.08}
    \resizebox{\columnwidth}{!}{%
    \begin{tabular}{lccc}
        \hline
        Variant
        & MAE (dB) $\downarrow$
        & High-Freq. MAE (dB) $\downarrow$
        & No-Harm Violation (\%) $\downarrow$ \\
        \hline
        XBase, $K=0$
        & 5.5354 & 13.4462 & -- \\
        Full BeamRMX, $K=2$
        & \best{3.4252} & \best{4.7323} & \best{10.18} \\
        \hline
        No Source BeamRM
        & 5.5363 & 15.5742 & 19.29 \\
        No Region Gate
        & 6.9513 & 7.1210 & 52.56 \\
        Random Support
        & \second{3.6979} & \second{5.8082} & \second{11.13} \\
        \hline
    \end{tabular}}
\end{table}
Table~\ref{tab:mechanism_analysis} reveals three complementary aspects of the evidence mechanism. Removing the source BeamRM content nearly eliminates the source-assisted gain: the MAE increases from 3.4252 to 5.5363~dB, essentially returning to the $K=0$ XBase level, while HF MAE deteriorates even further. The improvement therefore originates primarily from realized same-scene radio evidence rather than from the additional adapter structure or source-configuration metadata alone.

Useful evidence must also be transferred conservatively. Removing the spatial region gate increases MAE to 6.9513~dB, which is worse than the query-only XBase anchor, and raises the no-harm violation rate to 52.56\%. Thus, access to informative source BeamRMs alone does not guarantee a better target prediction: uncontrolled residual injection can introduce substantial negative transfer. This result directly supports the gated-residual design in Section~IV-C, which restricts where cross-configuration evidence is allowed to modify the frozen prediction.

Random support selection causes a smaller but consistent degradation. Random admissible supports still improve substantially over XBase, confirming that same-scene radio responses themselves constitute the dominant source of useful information. Nevertheless, the full method improves MAE from 3.6979 to 3.4252~dB and also achieves better HF accuracy and a lower no-harm violation rate. The evidence therefore indicates that support relevance affects how effectively same-scene information can be transferred, while the larger gains arise from the presence of realized source BeamRMs and the spatial control of their correction.

Intra-sector beam refinement utility:
We further evaluate whether these improvements in BeamRM reconstruction provide useful information for downstream beam management. Because the 64 target beams form a contiguous angular-sector codebook rather than a full-azimuth codebook, the task is formulated as intra-sector beam refinement after a candidate sector has been identified. Let $\mathcal{B}$ denote the target beam set. For scene $s$, the reference serving-power surface and the beam selected from the predicted BeamRMs are
\begin{equation}
\begin{aligned}
P^{\mathrm{serv}}_{s}(p)
&=
\max_{b\in\mathcal{B}}
\mathbf{Y}_{s,b}(p),\\
\widehat{b}_{s}(p)
&=
\arg\max_{b\in\mathcal{B}}
\widehat{\mathbf{Y}}_{s,b}(p).
\end{aligned}
\label{eq:beam_selection}
\end{equation}
The resulting beam-selection regret is defined as
\begin{equation}
R_s(p)
=
P^{\mathrm{serv}}_{s}(p)
-
\mathbf{Y}_{s,\widehat{b}_{s}(p)}(p),
\label{eq:beam_regret}
\end{equation}
which measures the received-power loss caused by selecting the beam favored by the predicted BeamRMs instead of the true best beam at location $p$. A regret of zero therefore indicates that the predicted BeamRMs select a ground-truth optimal beam. A Top-3 hit occurs when $\widehat{b}_{s}(p)$ belongs to the three beams with the highest ground-truth received powers at location $p$.

Serving-power MAE is computed over the spatial intersection in which all target beams are valid. The Top-3 hit rate and beam-selection regret are evaluated at locations whose reference serving power is no lower than $-110$~dBm; this threshold defines a reference-covered evaluation region rather than an operational coverage guarantee. To additionally assess the spatial localization of strong beam responses, the footprint intersection over union (IoU) is computed from the top 5\% power region of each target BeamRM and averaged over beams and scenes.

\begin{table}[!t]
    \centering
    \caption{Intra-sector beam refinement utility on 160 unseen scenes.}
    \label{tab:beam_utility}
    \footnotesize
    \setlength{\tabcolsep}{3.8pt}
    \renewcommand{\arraystretch}{1.08}
    \resizebox{\columnwidth}{!}{%
    \begin{tabular}{lcccccc}
        \hline
        \multirow{2}{*}{Metric}
        & \multicolumn{2}{c}{BeamCKM-Support}
        & \multicolumn{2}{c}{RadioDiff-Support}
        & \multicolumn{2}{c}{BeamRMX} \\
        \cline{2-7}
        & $K=0$ & $K=2$ & $K=0$ & $K=2$ & $K=0$ & $K=2$ \\
        \hline
        Serving MAE (dB) $\downarrow$
        & 6.735 & 5.525 & 6.458 & \second{4.053} & 6.071 & \best{3.612} \\

        Top-3 Hit (\%) $\uparrow$
        & 67.48 & 68.85 & 67.17 & 67.64 & \second{77.87} & \best{79.96} \\

        Mean Regret (dB) $\downarrow$
        & 4.148 & 3.671 & 4.367 & 3.528 & \second{3.439} & \best{2.873} \\

        P95 Regret (dB) $\downarrow$
        & 16.986 & 15.657 & 17.739 & \second{14.593} & 15.744 & \best{13.502} \\

        Footprint IoU $\uparrow$
        & 0.549 & 0.582 & 0.593 & \second{0.672} & 0.614 & \best{0.741} \\
        \hline
    \end{tabular}}
\end{table}

The source-assisted improvement remains evident at the beam management level. Relative to its query-only anchor, BeamRMX with $K=2$ reduces serving-power MAE by 40.5\% and mean beam-selection regret by 16.5\%, while also improving the Top-3 hit rate, the 95th-percentile (P95) beam-selection regret, and high response footprint overlap. It achieves the best result across all communication-oriented metrics in Table~\ref{tab:beam_utility}. These gains show that the additional radio evidence improves not only pointwise BeamRM reconstruction but also the relative ordering and spatial localization of candidate-beam responses, which are directly relevant to beam refinement.


The scene-level empirical distributions in Fig.~\ref{fig:communication_ecdf} further show that the improvement is not driven by only a small number of favorable environments. BeamRMX with $K=2$ shifts both the serving-power MAE and mean-regret distributions toward lower values relative to its $K=0$ anchor. A paired scene-level bootstrap analysis gives a mean serving-power MAE improvement of 2.460~dB, with a 95\% confidence interval of $[2.224,\,2.715]$~dB, and improvement is observed in all 160 test scenes. Together with the mechanism diagnostics, these results show that controlled use of same-scene cross-configuration evidence provides consistent gains from BeamRM reconstruction to downstream intra-sector beam refinement utility.

\section{Conclusion}
\label{sec:conclusion}

This paper investigated generalizable BeamRM prediction in beamformed wireless systems, where different beam and radio configurations produce distinct received power fields within the same propagation scene. BeamRMX addresses this one-scene--many-BeamRM setting by representing the target radiation state explicitly in the spatial domain and formulating BeamRM prediction as learning how the scene transforms this spatial radiation query into the corresponding received power field. This common spatial representation supports prediction across heterogeneous beam states and configurations, while XBase explicitly learns the interaction between radiation and scene geometry. Experiments under matched-domain, unseen-scene, and unseen-configuration settings consistently validate the effectiveness and generalization of this formulation. When same-scene radio evidence is available, the optional Evidence Adapter provides further refinement without target observations or test-time updates. The resulting BeamRMs also improve serving-power estimation and intra-sector beam refinement. Future work will investigate measurement-based validation and broader beam and array configurations.

\bibliography{ref}
\bibliographystyle{IEEEtran}

\end{document}